\documentclass[aip,pop,reprint,groupaddress,noshowpacs,noshowkeys]{revtex4-1}

\usepackage{amsmath,amssymb,mathrsfs}

\usepackage{graphicx}
\usepackage{dcolumn}
\usepackage{bm}
\usepackage{xcolor}
\usepackage[pdftex,
  hidelinks,
  colorlinks=true,
  linkcolor=black,
  anchorcolor=black,
  citecolor=black,
  urlcolor=black,
  pdfauthor= {Michael Glinsky},
  pdfsubject = {Mallat Scattering Transformation},
  pdfdisplaydoctitle,
  bookmarks=true,
  bookmarksopen=true]{hyperref}

\begin{document}

\title{Quantification of MagLIF morphology using the Mallat Scattering Transformation}

\author{Michael E. Glinsky}
\author{Thomas W. Moore}
\author{William E. Lewis}
\author{Matthew R. Weis}
\author{Christopher A. Jennings}
\author{David J. Ampleford}
\author{Patrick F. Knapp}
\author{Eric C. Harding}
\author{Matthew R. Gomez}
\author{Adam J. Harvey-Thompson}
\affiliation{Sandia National Laboratories, Albuquerque, New Mexico 87185 USA}


\begin{abstract}
    The morphology of the stagnated plasma resulting from Magnetized Liner Inertial Fusion (MagLIF) is measured by imaging the self-emission x-rays coming from the multi-keV plasma. Equivalent diagnostic response can be generated by integrated radiation-magnetohydrodynamic (rad-MHD) simulations from programs such as \texttt{HYDRA} and \texttt{GORGON}. There have been only limited quantitative ways to compare the image morphology, that is the texture, of simulations and experiments. We have developed a metric of image morphology based on the Mallat Scattering Transformation (MST), a transformation that has proved to be effective at distinguishing textures, sounds, and written characters. This metric is designed, demonstrated, and refined by classifying ensembles (\textit{i.e.}, classes) of synthetic stagnation images, and by regressing an ensemble of synthetic stagnation images to the morphology (\textit{i.e.}, model) parameters used to generate the synthetic images.  We use this metric to quantitatively compare simulations to experimental images, experimental images to each other, and to estimate the morphological parameters of the experimental images with uncertainty.  This coordinate space has proved very adept at doing a sophisticated relative background subtraction in the MST space.  This was needed to compare the experimental self-emission images to the rad-MHD simulation images.
\end{abstract}



\maketitle

\section{\label{sec:intro}Introduction}
Magnetized Liner Inertial Fusion (MagLIF) is a magneto-inertial fusion concept currently being explored at Sandia's Z Pulsed Power Facility.\citep{Slutz2010,Awe2013,Gomez2014,McBride2012}  MagLIF produces thermonuclear fusion conditions by driving mega-amps of current through a low-Z conducting liner. The subsequent implosion of the liner containing a preheated and pre-magnetized fuel of deuterium or deuterium-tritium compresses and heats the system, creating a plasma with fusion relevant conditions.

\begin{figure}[ht]
\includegraphics[width=\columnwidth]{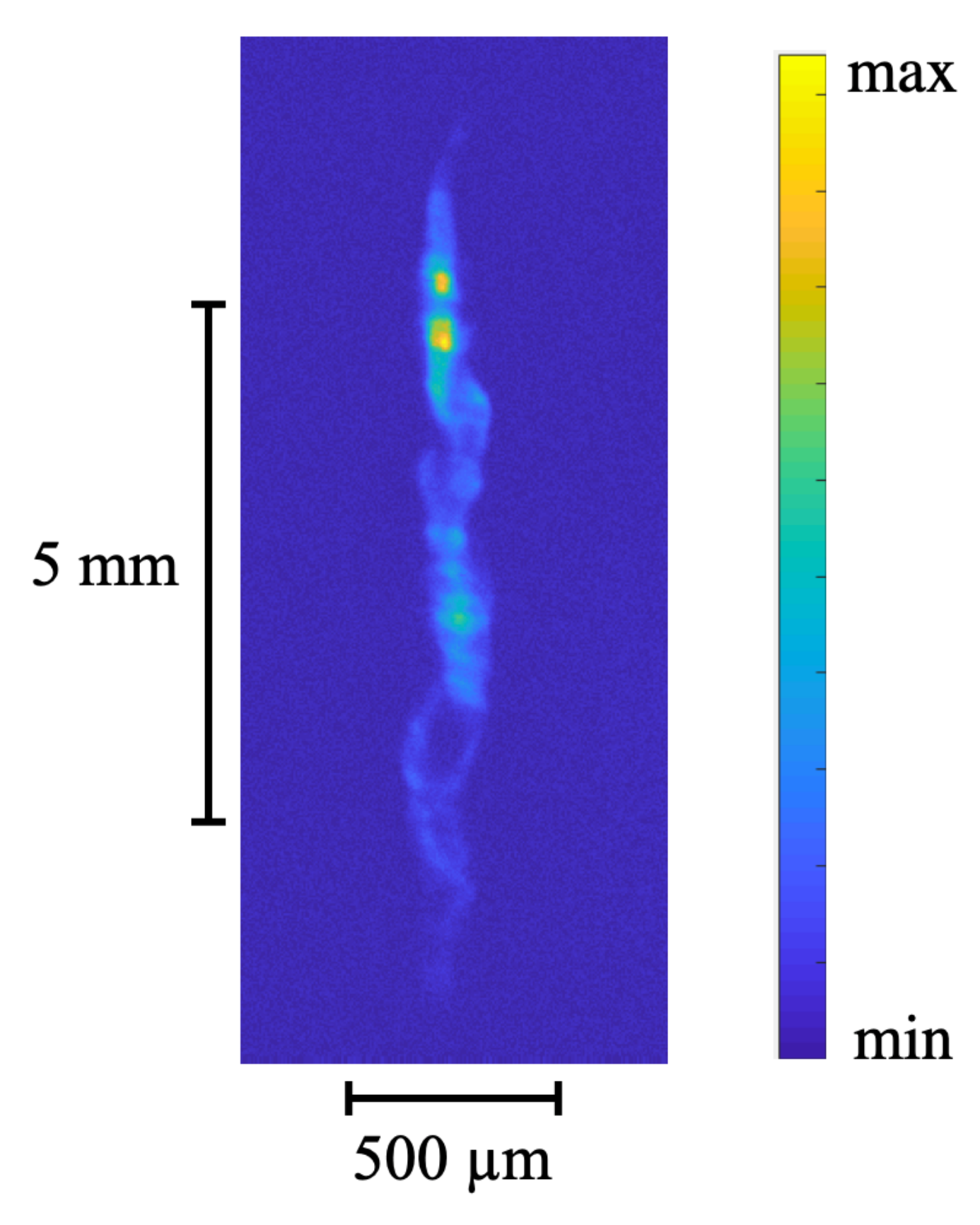}
\caption{\label{fig:helix} Self emission x-ray image of fuel plasma stagnation showing double-helix structure (experiment z3236).  Axial direction is vertical.  Radial direction is horizontal and exaggerated.}
\end{figure}

Developing a detailed understanding of how experimental parameters such as axial pre-magnetization, preheat, and liner design mitigate losses and affect performance, as well as evolution of the plasma, is a crucial and ongoing step towards realizing the full potential of MagLIF. To this end, time resolved radiography of the imploding liner, as well as self-emission x-rays from the fuel plasma at stagnation (where thermal pressure of the fuel plasma stalls the liner implosion), have been used to study the evolution of the plasma and its structure at peak fusion conditions.  For example, \citet{Awe2013} observed an unexpected feature in radiographs of an axial magnetized imploding liner -- a multi-helix structure not observed in liners that were not axially premagnetized.  Additionally, axially bifurcated double helical strands have been observed in the stagnating fuel plasma columns, captured by self-emission x-ray image diagnostics.  See Fig.~\ref{fig:helix} for an example image of the x-ray self-emission from the stagnated fuel plasma.  Details of the helical structure vary, such as if there are one or two strands, and may not be resolved in some images since the resolution of the x-ray imager has just recently been improved.

The underlying physics linking the multi-helix structure of the imploding liner to the bifurcated double-helices in the stagnated plasma is as of yet unknown. One working hypothesis is that a helical magnetic Rayleigh-Taylor instability \citep{Seyler2018} (MRT) seeded on the outside liner surface may grow large enough to feed-through the liner to seed perturbations on the liner interior. It is thought that these interior perturbations may imprint the double helical structure on the plasma. It has been experimentally demonstrated that the helical structure is dependent on the aspect ratio of the liner (AR $\equiv$ liner's initial outer radius$ / $liner's initial wall thickness).  Recent experiments using liners with no dielectric coating appear to demonstrate that for increasing AR, the stagnation column helical radius increases while helical wavelength decreases, which is consistent with MRT feed-through from the liner’s outer surface.\citep{Ampleford2019}  Dielectric coatings are sometimes used on the outside of the liner to suppress the electro-thermal instability which can strongly seed the MRT.\citep{peterson2012electrothermal,peterson2014electrothermal}   There is another, less developed, working hypothesis that this double helical structure might be an emergent structure of the nonlinear evolution of the MRT that is controlled by conserved magnetic and cross  helicities (where cross helicity\citep{perez2009role, glinsky2019helicity} is the cross-correlation between the fluid velocity and magnetic field averaged over an ensemble of random motions) that are injected into the liner.  The large scale self organization would be the result of a Taylor relaxation,\citep{taylor1986relaxation} that is an energy minimization under the constraints of the topologically conserved helicities.  This is supported by the inverse turbulent cascade in the liner structure seen by \citet{yager2018} on ultra-thin foils driven at less than 1 MA.  However, such inferences remain weak due to the fact that, to date, there has been no systematic way to quantitatively compare stagnation morphology experiment-to-experiment or experiment-to-simulation while accounting for the uncertainty in characterizing features such as the helical wavelength and radius.

In this work, we develop a method which enables such a comparison by applying a cutting edge Machine Learning (ML) algorithm in image classification known as the Mallat Scattering Transform (MST).\citep{Mallat2012,Bruna2013}  Specifically, we are able to use the MST as a quantitative metric of morphology to compare stagnation images, and as a metric to infer morphological features with uncertainty via a regression. We start with two sections that are a thorough description of the theoretical methods and details of the technical approaches that are essential in enabling other researchers to apply this approach to their data.  They are quite dense and can be skimmed paying particular attention to the figures after reading the main take aways at the beginning of each of these sections. In Sec.~\ref{sec:theory}, we supply the required theory for the MST, show its connection to Deep Learning, and describe its relationship to causal physics. Section \ref{sec:ml_pipeline} describes the synthetic model used to parametrize the double helix morphology.  We then discuss the design of the image morphology metric based on the MST.  This metric is then tested in two ways.  The first is via a classification of ensembles of synthetic stagnation images, and the second is via performance of a full machine learning pipeline that quantifies the morphological parameters of the stagnation images with uncertainty via regression.  Section~\ref{sec:ml_pipeline} concludes with a verification of the metric design.  Section~\ref{sec:results} demonstrates the application of the metric of image morphology in quantitatively comparing simulation and experiment, as well as a direct extraction of the morphological parameters with uncertainty from experimental images. We highlight the viability of the method to differentiate between plasmas produced from different experimental designs, and the use of the MST to do a sophisticated background subtraction to enable comparison of experiments to simulations.

\section{\label{sec:theory}Mallat Scattering Transform}
The MST is an iterative transformation that consists of convolutions with a localized oscillatory wavelet dilated to various scales that are nonlinearly rectified with a modulus operation then iteratively repeated.  After each iteration the transformation is averaged over local patches and output.  This gives a nonlinear mapping of the spacial features of the image to its scale features, including multiple scale correlations.  This is a particular form of what is called a Convolutional Neural Network (CNN), a form of deep learning.  It has a very specific, predetermined form with only a couple design parameters that need to be determined.  This allows it to be trained on very small datasets.  It also has very deep connections to physics that makes it a very compact and efficacious encoding. It respects the constraints of the physics such as causality, advective continuity, topological conservations, and more traditional conservations generated by group symmetries.

\subsection{\label{sec:DL_MST_theory}Deep learning based definition of MST}
Recently, the use of deep learning methods, combined with availability of large labeled data sets, has enabled a revolution in the field of image classification and analysis.  Particularly, CNNs have gained widespread popularity for image analysis problems, such as classification,\citep{LeCun1989} segmentation,\citep{Ning2005} and even image generation.\citep{Goodfellow2014}  The ubiquity of this approach is largely based on the ability of CNNs to learn convolutional filters which compute features that are approximately invariant to irrelevant symmetries present in the task (\textit{e.g.} translation or rotational symmetries).\citep{LeCun2010} 

However CNNs require significant expertise to navigate a seemingly arbitrary design space (\textit{e.g.}, number of nodes and layers) and require considerable computing resources to train, even when using transfer learning.\citep{goodfellow2016deep}  Additionally, their {\textit{black box}} nature make CNNs a less attractive framework for scientific applications to bridge the gap between causation and correlation. Alternative kernel classifiers such as the probabilistic neural network, are based on the Euclidean distance between image features (\textit{e.g.}, pixel information), which is easily broken by transformations, rotations and scaling. At the same time, familiar translation invariant feature representations such as the Fourier transform modulus are unstable to deformations (that is not Lipschitz continuous). The wavelet transformation on the other hand, is Lipschitz continuous to deformation, but is not translation invariant.\citep{Bruna2013}  By combining local translation invariance and Lipschitz continuity to deformations in a fixed weight convolutional network, the MST addresses many of the concerns that arise in deep learning.\citep{Mallat2012,Bruna2013}

\begin{figure}[ht]
\includegraphics[width=\columnwidth]{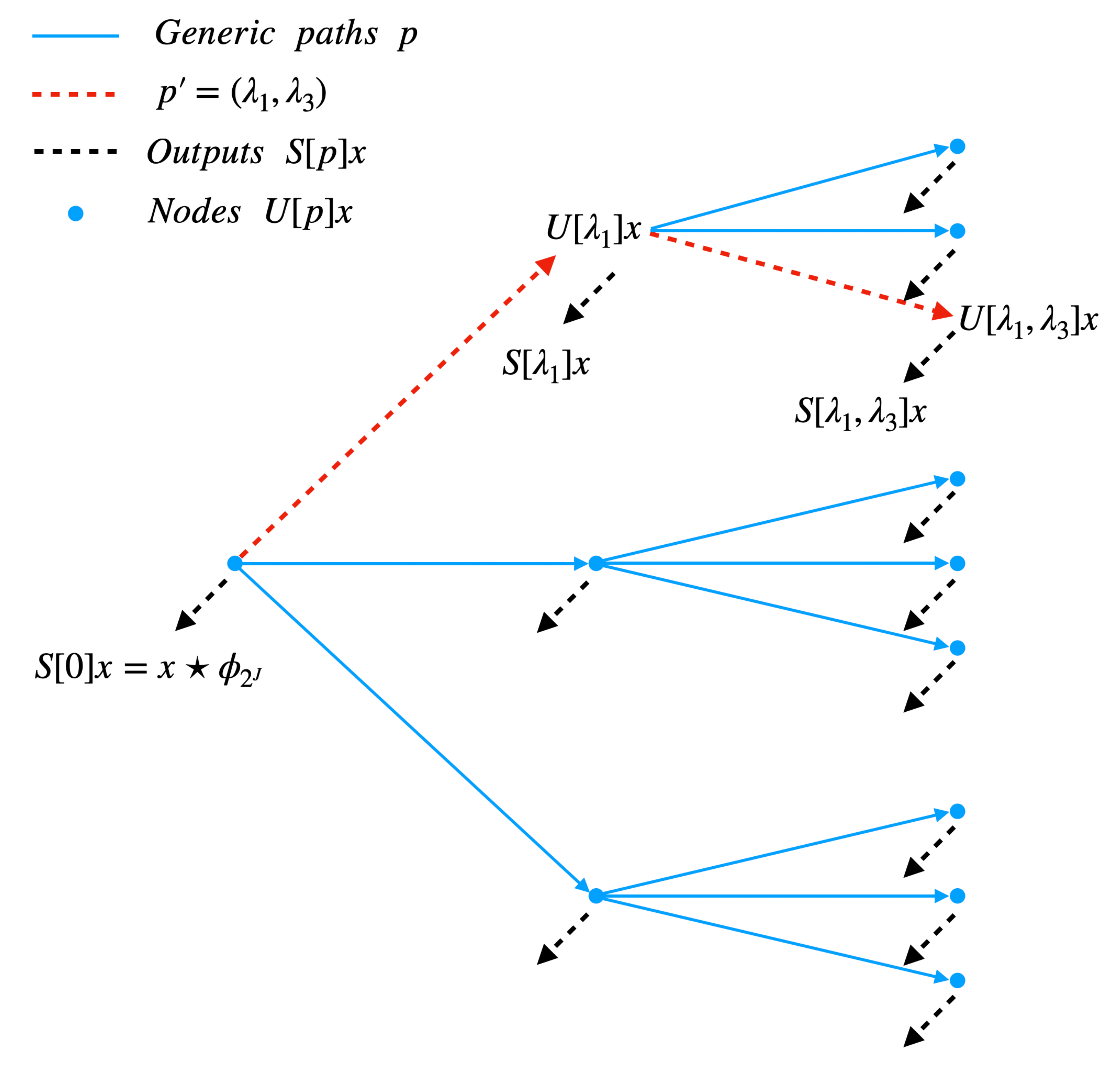}
\caption{\label{fig:mst} The MST may be thought of as a convolutional network with fixed weights. The above network could represent for example a 1D MST with 3 scales, and no rotations (1D case). The network outputs MST coefficients averaged by a Father Wavelet along each path, $S[p]x$. Each node of the network is the set of scattering coefficients before being window averaged by the Father Wavelet, $U[p]x$. The operator $\tilde{W}$ of Eq.~\eqref{eqn:MST} expands the network below a given node at then at the end of a path, $p$.}
\end{figure}

The MST consists of compositions of wavelet transformations\citep{mallat1999wavelet} coupled with modulus and non-linear smoothing operators which form a deep convolutional network. Unlike traditional deep convolutional neural networks, the filters in the MST are prescribed rather than learned. In fact the deep convolutional network of the MST has been shown to outperform CNNs for image classification tasks over a broad range of training sample sizes.\citep{Bruna2013} This is most significant when the amount of training samples is considerably limited,\citep{Bruna2013} which is often the case with experimental data. Additional benefits of the MST framework over CNNs come in the form of intelligible design -- for example, the depth of an MST network is bound by a signal's energy propagation through the network, whereas the depth of a CNN is seemingly arbitrary.

\begin{figure}[h]
\includegraphics[width=\columnwidth]{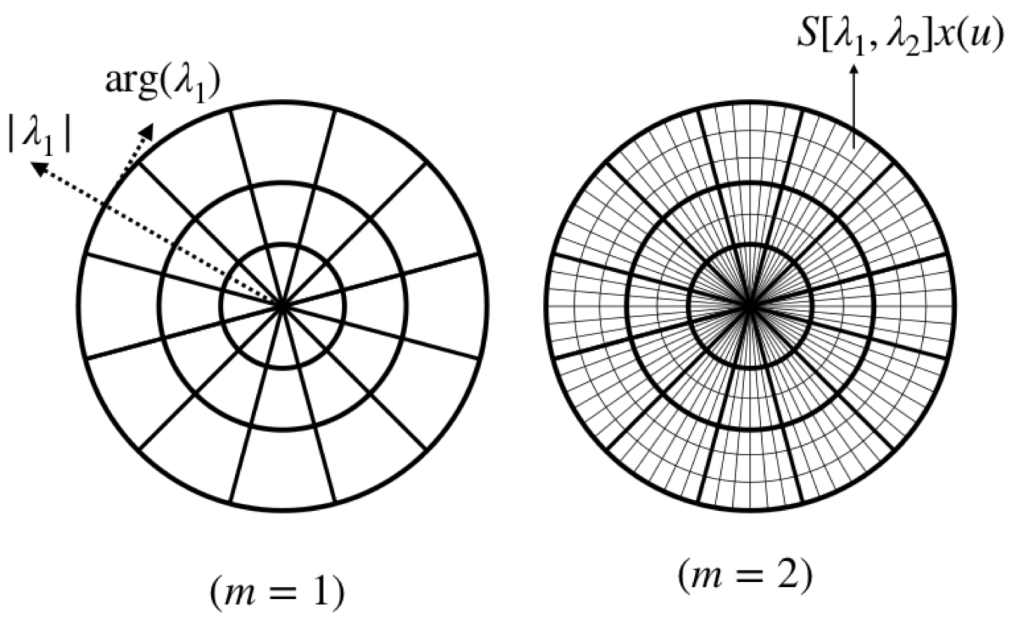}
\caption{\label{fig:logpolar} Coefficients produced by applying MST to 2D images in this work will be displayed on radial plots as shown. Bins are created according to scale (radial positioning, $|\lambda_m|$, and rotation, $\text{arg}(\lambda_m)$) with magnitude (color scale, not shown) representing the size of the coefficient at that scale and rotation.}
\end{figure}

The two-dimensional MST uses a set of convolutional filters which are calculated from a Mother Wavelet $\psi$ by applying a rotation $r$ and scaling by $2^j$:
\begin{equation}
\label{eqn:wavelet}
\psi_{\lambda} = 2^{-2j} \psi (2^{-j} r^{-j} u),
\end{equation}
where $\lambda=2^{-j} r$ and $u$ is the spatial position. Let the wavelet transformation of image $x(u)$ be given by $x \star \psi_\lambda$. Given that the spatial resolution is retained in a wavelet transform, this process can be iterated upon, such that the propagated signal along path $p = (\lambda_1, \lambda_2,\dots,\lambda_m)$ is given by:
\begin{eqnarray}
	U[p]x &= U[\lambda_m] = U[\lambda_2] \cdots  U[\lambda_m]x \nonumber \\
	 &= | || x \star \psi_{\lambda_1} | \star \psi_{\lambda_2} | \cdots | \star \psi_{\lambda_m} | \label{eqn:wavelet_of_wavelet}
\end{eqnarray}
where the modulus removes the complex phase from the propagated signal. However, the wavelet coefficients are not invariant to translation, but rather translation covariant. Introducing the Father Wavelet (\textit{i.e.}, a spatial window function), $\phi_{2^J}(u)=2^{-2J}\phi(2^{-J}u)$, allows an average pooling operation to be performed by convolution $U[p]x \star \phi_{2^J}(u)$. This operation collapses the spatial dependence of the wavelet coefficients while retaining the dominant amplitude $U[p]$ at each scale. This results in an effective translation invariance assuming that a given translation is much smaller than the window scale, $2^J$. The windowed scattering transformation is thus given by:
\begin{eqnarray}
	S[p]x(u) &=&U[p]x \star \phi_{2^J}(u) \nonumber \\
	&=&| || x \star \psi_{\lambda_1} | \star \psi_{\lambda_2} | \cdots | \star \psi_{\lambda_m} | \star  \phi_{2^J}(u). \label{eqn:windowed_scattering}
\end{eqnarray}
Now, we may define an operator $\tilde{W}$ which acts upon the non-windowed scattering $U[p]x$ producing
\begin{equation}
	\tilde{W} U[p] x = \{S[p]x, U[p + \lambda]x \}_{\lambda \in \mathcal{P}} \label{eqn:MST}.
\end{equation}
$\tilde{W}$ will produce the output scattering coefficient at the current layer for the given path $p$, and will move to the next layer along the path $p+\lambda$ as demonstrated in Fig.~\ref{fig:mst}.  With Eqns. \eqref{eqn:wavelet_of_wavelet} and \eqref{eqn:windowed_scattering}, we arrive at a deep scattering convolutional network $\tilde{W}$ in Eq.~\eqref{eqn:MST} with $m$ layers. For 2-D signals (images), the MST coefficients are visualized via log polar plots as depicted in Fig.~\ref{fig:logpolar}.

The MST forms a nonlinear mapping from an image's spatial features to its scale features. This mapping is Lipschitz continuous to deformation, meaning that small deformations of the image result in small deformations of the Mallat scattering coefficients. Since we will be concerned with discovering morphology parameters of stagnation column images such as helical wavelength, the MST provides a convenient basis as compared to, for example, a Fourier transform which is not Lipschitz continuous to deformations.  The first order MST, $m=1$, can be viewed as an optimal ``local'' Fourier transform. The reader is referred to \citet{Mallat2012} and \citet{Bruna2013} for more details regarding the mathematical properties of MST, such as being a unitary transformation, and having a scale ordering of the path, $p$.

\subsection{\label{sec:physics_MST_theory}Physical foundation of MST}
In the previous Sec.~\ref{sec:DL_MST_theory}, we developed the MST as a deep convolutional network with a very specific form. There was no physical reason given (other than desiring the mathematical properties of Lipschitz continuity and stationarity) for the design choices such as:  the use of iterative convolution with a Mother Wavelet, the use of the modulus as an activation function between layers, and the final pooling using convolution with the Father Wavelet. There was also no reason given for: the sparseness of the MST, the need for only the first and second order MST, and the efficacy of the MST as a representation for the Machine Learning of physical systems. As it turns out, there are deep physical foundations for these design choices that explain its compactness and efficacy.  These connections were briefly mentioned in \citet{Mallat2012} and are alluded to by the use of the word ``Scattering'' in the name of the transformation.

Important properties that connect the MST to causal dynamics have been noted in the previous section. Fundamental to these connections is the fact that physical dynamics, whether it be fluid dynamics, classical mechanics, or quantum field theory is built upon advection (\textit{i.e.}, deformation) by a vector  field, which is also how physical symmetries are generated. This is why having Lipschitz continuity is paramount. Imposing this constraint on the transformation limits the representation to physically realizable systems with the proper symmetries.  Furthermore, the construction of the MST also leads to the properties that the transformation is unitary (expressing that probability can not be created or destroyed), and that the path is scale ordered (expressing that the system is causal).  Furthermore, the convolution by the Father Wavelet and the use of the modulus can be viewed as an expectation value operator and the evaluation of Gaussian integrals via the method of stationary phase, respectively.

Another seemingly arbitrary choice is the truncation of the MST expansion at second order.   A physical system that encodes finite information is fully identified by the first and second order MST.  This is because of a statistical realizability theorem\citep{krommes2002} -- either the distribution stops at second order or it must continue to all orders. Since the dynamical information is finite, the distribution must stop at second order.  Practically, it is found that there is little signal energy in the MST of third order and higher, and that there is almost no improvement in the classification or regression  performance by including the third order MST.

Finally, it is worth noting that, in the context of images, the MST is encoding the static scale structure in the first order transform, and the relationship between structures of different scales in the second order transform. This scale-to-scale correlation is essentially a two-point correlation function between different locations in the image, and the first order transform is the single-point correlation function (essentially a local Fourier transform).  In the context of dynamical systems, this is analogous to the single-particle and two-particle distribution functions in the Mayer Cluster expansion of classical kinetic theory and the equivalent constructs of quantum field theory. These quantities encode all the dynamics of the system, meaning that the MST has a profound connection to the underlying physical dynamics of the system, and is a compact, that is sparse, encoding of the dynamics.  Another way of looking at this is that the physics of the system is encoded in the two scale correlation function -- the kinetic or quantum transition rates.  These transition rates determine how the first order transform evolves.  This encoding of the physics in the second order transform is not surprising.  In plasmas the electrostatic force is encoded in Debye shielding which is a two-point correlation.  Significant theoretical and numerical work is well underway to support these physical foundations of the transformation.\citep{glinsky.et.al.19, glinsky.11}

\section{\label{sec:ml_pipeline}Synthetic model, classification and regression}
An analytic, synthetic model is constructed based on 11 model parameters that describe a 2D image of a double helical stagnation.  There are also 6 stochastic parameters that represent the stochastic nature of the stagnation image and the experimental measurement noise.  This model is used to generate data sets to design the metric based on the MST, train a classifier, and to develop a regression for the model parameters given an image.  The design of the metric consists of choosing:  the patch size over which the transformations will be done, which patches to include in the analysis, the prior distributions of the features (whether it is normal or log-normal, essentially whether the log of the features should be taken), how the features should be normalized, the angular resolution of the MST, and the maximum order of the MST to use.  These choices are optimized and validated, and the efficacy of the metric demonstrated by the quantified performance of the classification and regression.  The metric can be used directly in the analysis, and the regression can be used to estimate the model parameters (of the synthetic model) with uncertainty given an experimental image.

\subsection{Synthetic double helix model}
In order to quantify the morphology of the MagLIF stagnation column, a model with well defined parameters is needed to act as a surrogate for the x-ray self-emission diagnostic images.  This model must capture the essential features of the stagnation such as its multi-helical nature, finite axial extent and axial bifurcations. For this purpose, we have constructed a synthetic model complete with 11 descriptive model parameters that capture some features of a fundamentally 3D stagnating plasma projected into a 2D image along with 6 stochastic parameters to represent the {\textit{natural}} experimental variation and signal noise inherent in the x-ray diagnostics fielded on Z.  

Analytically, the synthetic model consists of superimposed ``radial'' and axial Gaussians over a pair of axial $\cos^2$ waves.  Here, the radial position projected onto the image will be given by $r$, and the axial position of the image will be given by $z$. The model may be specified by the composition of the following functions:
\begin{eqnarray}
\label{eqn:s}
s(z) &=& \theta_6 \cos^2(\theta_7*\theta_3*z+\zeta_5) \nonumber\\ 
&+&\theta_9 \cos^2(\theta_{10}*\theta_3*z+\zeta_6),
\end{eqnarray}
\begin{equation}
\label{eqn:r0}
r_{0,j}(z)= (-1)^{1+\delta_{j,2}}\theta_8 + \theta_5 \sin(\theta_3*z + \zeta_4 + \delta_{j,2} \theta_{11}),
\end{equation}
\begin{equation}
\label{eqn:g}
g_j(r,r_{0,j}(z)) = \frac{1}{\theta_1 \sqrt{2\pi}} \exp\Big\{\frac{-(r-r_{0,j}(z))^2}{2 \theta_1^2}\Big\},
\end{equation}
\begin{equation}
\label{eqn:ell}
\ell(z) = \frac{\zeta_3}{\theta_2 \sqrt{2\pi}} \exp\left\{ -\left( \frac{z^2}{2 \theta_2^2} \right)^{\theta_4} \right\},
\end{equation}
and
\begin{eqnarray}
\label{eqn:h}
h(r,z) &=& A \sum_{j=1}^2\Big[ (1+s(z)) \; g_j(r,r_{0,j}(z)) \; \ell(z) \nonumber\\
&\times& (1-\zeta_2 U(0,1)) + \zeta_1U(0,1)\Big],
\end{eqnarray}
where $h(r,z)$ is the final composition used to generate double helix images, $\ell(z)$ is the axial envelope, $g_j$ is the Gaussian envelope of the helical strand, $r_{0,j}(z)$ is the center of the helical strand, $s(z)$ is axial bifurcations of the helical strands, $U(0,1)$ is a uniformly distributed random number on $[0,1]$, $j \in \{1,2\}$ is the strand index, $\delta_{i,j}$ is the Kronecker delta function, $A$ is a constant used to normalize the max value of $h(r,z)$ to unity, and $(\theta_i, \zeta_i)$ will be described. 

\begin{figure}[ht]
\includegraphics[width=\columnwidth]{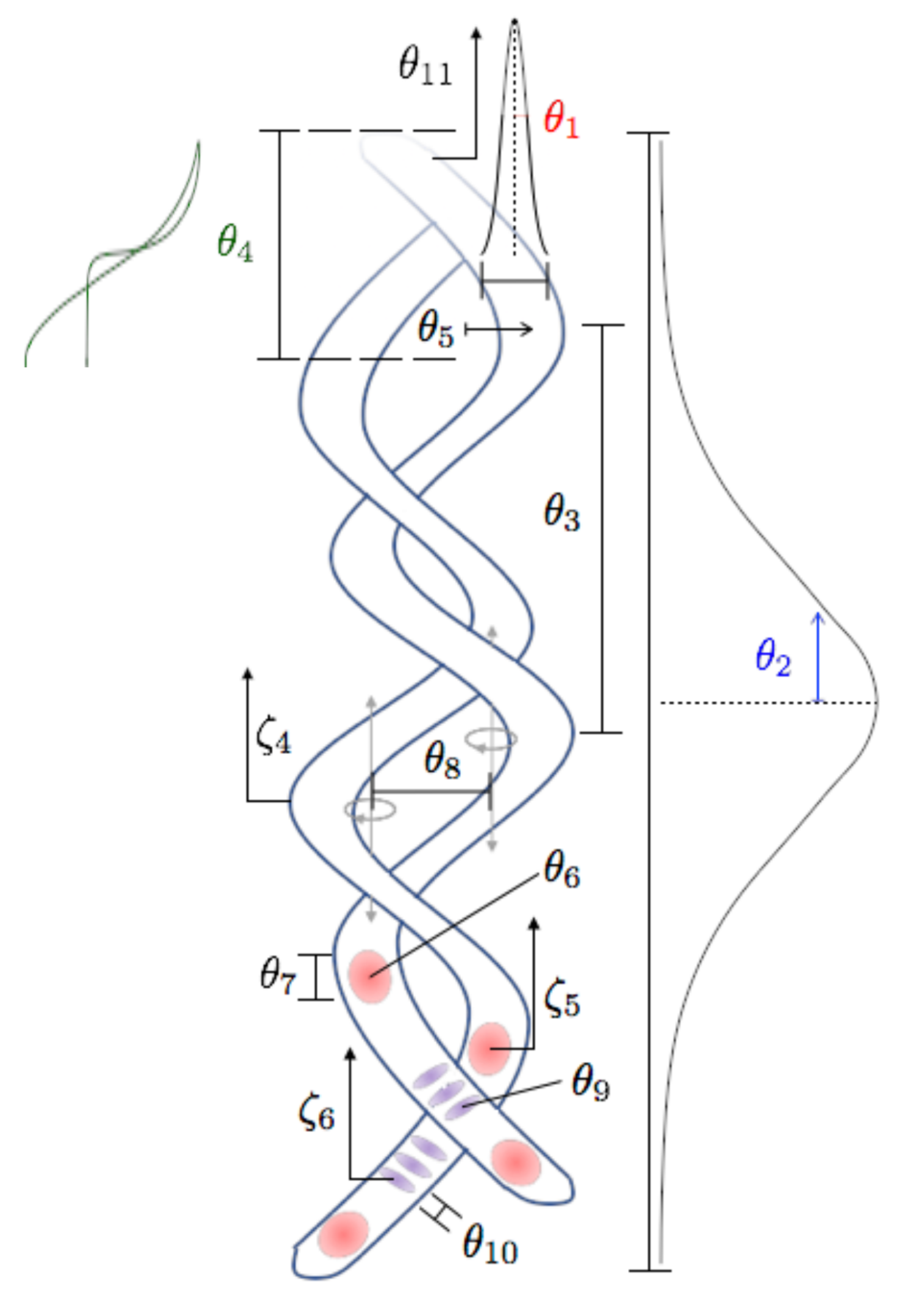}
\caption{\label{fig:syn} Synthetic Stagnation Model  (see Table~\ref{tab:param}).}
\end{figure}

The $\theta_i$ and $\zeta_i$ parameters are depicted in Fig.~\ref{fig:syn} and summarized in Table \ref{tab:param}. Their interpretations are: $\theta_1$ is the standard deviation of the radial Gaussian or strand thickness,   $\theta_2$ is the standard deviation of the axial Gaussian or strand length, $\theta_3$ is the helical wavelength or wavenumber, $\theta_4$ is the order of the axial super Gaussian or how quickly the strand ends, $\theta_5$ is the amplitude of radial perturbations or radius of the helix, $\theta_6$ is the amplitude of the large wavelength or low frequency axial brightness perturbations, $\theta_7$ is the wavelength or mode number of the large wavelength or low frequency axial brightness perturbations, $\theta_8$ is the strand separation, $\theta_9$ is the amplitude of the small wavelength or high frequency axial brightness perturbations, $\theta_{10}$ is the wavelength or mode number of the small wavelength or high frequency axial brightness perturbations, and $\theta_{11}$ is the relative strand phase; $\zeta_1$ is the background noise, $\zeta_2$ is the signal noise, $\zeta_3$ is the amplitude of the signal, $\zeta_4$ is the radial perturbation phase shift, $\zeta_5$ is the phase shift of the large wavelength or low frequency axial brightness perturbations, and $\zeta_6$ is the phase shift of the small wavelength or high frequency axial brightness perturbations.

\begin{table}
\caption{\label{tab:param}Synthetic model $\theta_i$ and stochastic $\zeta_i$ parameters (see Fig.~\ref{fig:syn}).}
\begin{tabular}{l}
\textbf{Model Parameters}\\
\hline
$\theta_1$ = strand thickness\\
$\theta_2$ = strand length\\
$\theta_3$ = helical wavenumber\\
$\theta_4$ = order of axial super Gaussian\\
$\theta_5$ = radius of the helix\\
$\theta_6$ = amplitude of low frequency axial brightness\\
perturbations\\
$\theta_7$ = mode number of low frequency axial brightness\\
perturbations\\
$\theta_8$ = strand separation\\
$\theta_9$ = amplitude of high frequency axial brightness\\
perturbations\\
$\theta_{10}$ = mode number of high frequency axial brightness\\
perturbations\\
$\theta_{11}$ = relative strand phase\\
\\
\textbf{Stochastic Parameters}\\
\hline
$\zeta_1$ = background noise\\
$\zeta_2$ = signal noise\\
$\zeta_3$ = amplitude of signal\\
$\zeta_4$ = radial perturbation phase shift\\
$\zeta_5$ = low frequency axial brightness perturbations phase\\shift\\
$\zeta_6$ = high frequency axial brightness perturbations\\ phase shift
\end{tabular}
\end{table}

\subsection{Metric design}
Given an image of the MagLIF stagnation column, we will calculate the MST coefficients as, what is commonly called in the ML literature, features of the image.  We will then use these features as a metric on the space of the images.  That is, the distance between two images will be calculated as the square root of the sum of the squares of the differences between the features of the two images.  There are several design decisions in calculation of the features.

A majority of the design decisions including gridding, maximum MST order, variable transformations, normalization, and scale resolution are inherited from \citet{Bruna2013}'s use of deep scattering transformation networks for handwritten digit recognition from the MNIST database of handwritten digits.  With this being said, these design decisions were verified by examining the effect on the regression performance of alternative design decisions.  See Sec.~\ref{sec:metric_verify}, for the verification.

\begin{figure*}[ht]
\includegraphics[width=2\columnwidth]{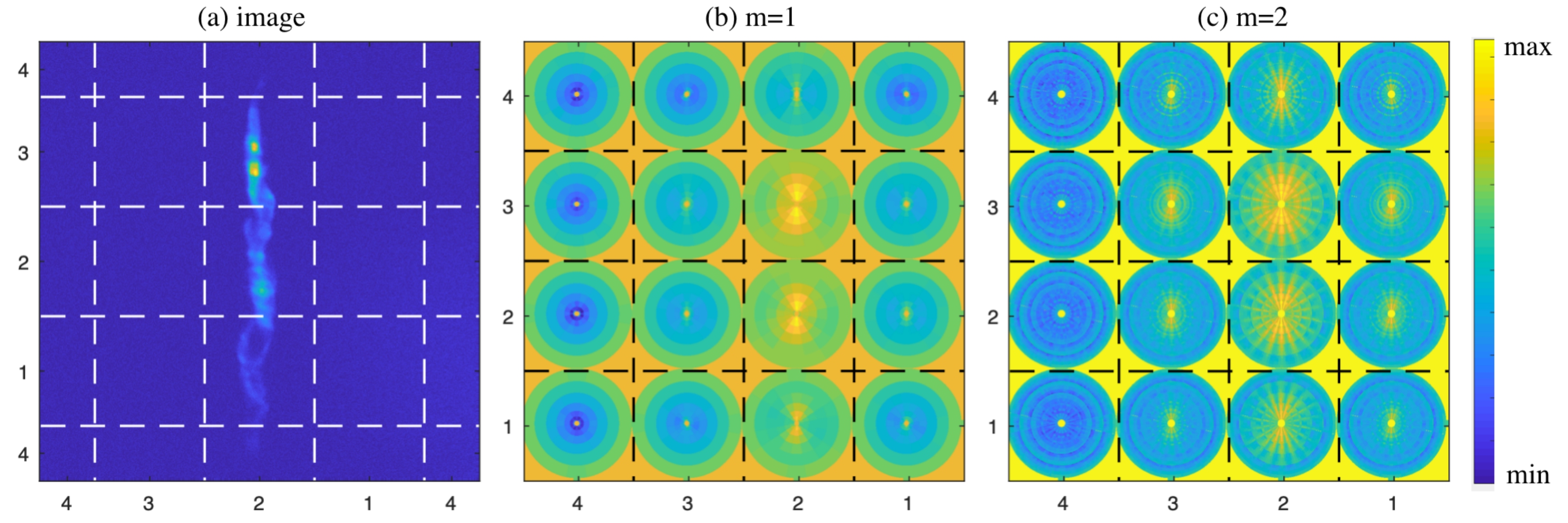}
\caption{\label{fig:grid} Gridded MST: (a, left) self-emission image of experiment z3236, (b, center) first and (c, right) second order MST coefficients.}
\end{figure*}

We now discuss how features were engineered. First, the reader may note from Eq.~\eqref{eqn:windowed_scattering} that we must evaluate the scattering coefficients at points $u$ in our image. Now, due to the assumption that the statistics given by the MST are stationary, that is spatially invariant, below the Father Wavelet window size; if we were to evaluate $S[p]x(u)$ at all points $u$, one would obtain very redundant information. As a result, it is wise to subsample $u$. This is achieved by translating the spatial window by intervals of $2^J$ such that $G_\#=N2^{-J}$, where $N$ is the symmetric pixel count and $G_\#$ is symmetric grid number. This subsampling forces each image to be segmented into a $G_\#\times G_\#$-grid.\citep{Bruna2013}  We work with images of pixel size $512\times512$, and set $J=7$ giving $G_\#=4$.  We now have a design parameter to choose, $J$, which determines the size of the sub-image, $2^J \times 2^J$, over which the transform will be calculated. This was chosen based on the position, size and characteristics of our double helix (see Fig.~\ref{fig:grid}) and was found to give good regression performance as discussed later. We note however, that a more rigorous procedure to select $J$ would be to select the $J$ which gives maximum cross validated classifier or regression performance.  With this being said, our eyes are very good at recognizing the dynamical space scale of the physics.  The size of the Father Wavelet, $2^J$, should be of this scale.  If the size is too small, the MST will be not be calculated over the largest area possible and will therefore have more noise and not contain as much statistical information.  If the size is too large, the assumption of stationarity will be violated leading to a blurring of the statistics and a resulting loss of information.  It is therefore expected that there will be an optimal size that could be determined by the aforementioned $J$ cross validation optimization.

This division of the image into sub-images, or patches, does not increase or decrease the resolution of the image.  As the image is divided into more and smaller patches, the number of pixels per sub-image decreases to compensate for the increased number of patches.  This also will decrease the number of MST coefficients per patch, since coefficients for scales larger than the patch size can not be calculated.

An added benefit to gridding the images, is data reduction via patch selection. From Fig.~\ref{fig:grid}(a) it is apparent that most of the image is background noise. This is echoed in the MST coefficient space. Since our double helix is confined to column $2$, essentially all of the unique information is contained within the MST coefficients evaluated on the four patches in column $2$, so that the other columns may be dropped.  This is done after the MST, not before, to mitigate boundary effects and because of technical issues in taking the MST that require the domain to be square.  We also only calculate the MST to second order, $m=2$.  Before computing the MST on our gridded image, we must apply boundary conditions for the convolution. There are many reasonable choices,  such as periodic, zero-padded, and mirrored. We chose to use a mirror boundary condition.  This minimized the influence of the boundary, while making the minimum assumption about the signal outside of the domain.

The final step in engineering features for a machine learning algorithm is to perform an appropriate scaling of the input features. This is a common practice in statistical learning, and many different scaling transformations and dimensionality reduction methods are reasonable. Here, we apply a $\log_{10}$ scaling to our scattering coefficients and model parameters (with the exception of $\theta_{11}$ which is a phase shift) used in training the classifier. This choice was made to decrease the dynamic range of the MST coefficients, since before the transformation the MST coefficients were dominated by only a few coefficients.

\subsection{Classification model}
\begin{figure*}[ht]
\includegraphics[width=2\columnwidth]{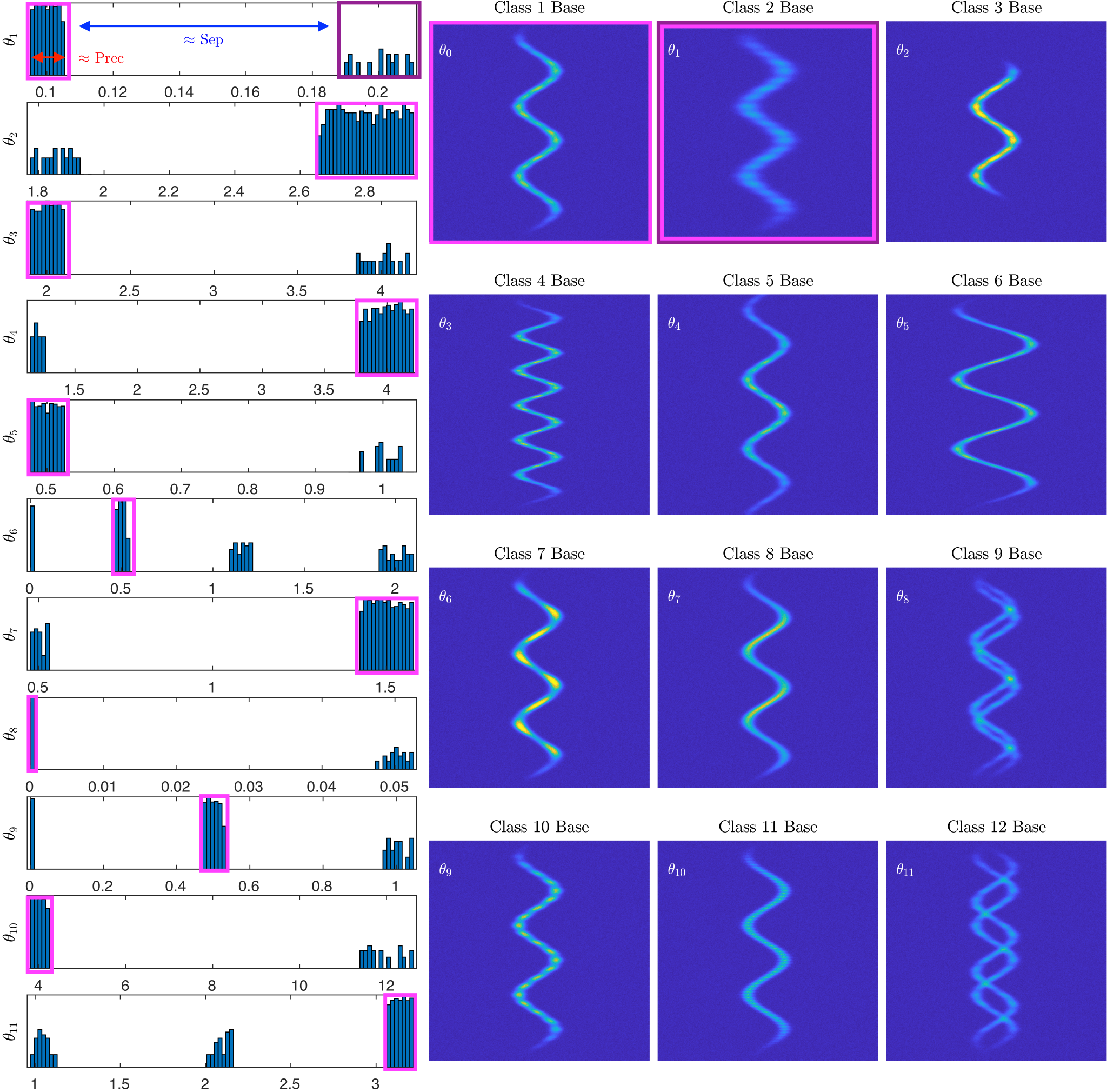}
\caption{\label{fig:classes} Classification Training Set construction. Parameter distributions are shown at left while the base classes are represented on the right.  Shown is the class separation between class 1 and 2 as the blue line labeled ``Sep'', and the class precision of class 1 as the red line labeled ``Prec''.   Units for all $\theta_i$ are arbitrary, but consistent with those to be shown in the cross plots of Fig.~\ref{fig:mstr_performance_param}.  Ensemble of synthetic stagnation images for classification (shown via animation in this link to a \href{https://youtu.be/PG4-rQ1tUyw}{Multimedia View}).  Shown in this animation are the base case (left), the base case for the class (middle), and the individual members of the ensemble (right).}
\end{figure*}
Studying the ability of the MST to distinguish between different classes of helical morphology will quantify the performance of the MST as a metric of image morphology and will provide reassurance that the regression problem is well-posed. Additionally, it provides access to more easily interpretable results (\textit{e.g.}, classification accuracy as opposed to $R^2$). By considering the classification problem, we are also able to closely follow the approach using the MST for MNIST handwritten digit recognition in \citet{Bruna2013}.

We approach the problem by synthesizing 12 stagnation image classes -- 11 distinct parameter constrained classes constructed from systematic modifications to the synthetic model parameter distributions from a single base class. Each of the distinct parameter classes has a definitive associated synthetic model parameter. For a given parameter class, the distribution of its associated synthetic model parameter is translated some separation from its corresponding base class distribution. This process is repeated for each of the 11 distinct parameter classes (see Fig.~\ref{fig:classes}). For the classification problem, we generate 340 images. We use $50\%$ of this data set as the training set to train an affine classifier, while the remaining $50\%$ is separated out as the test set to be used for characterizing the trained classifier.

Finally, we apply the classification algorithm. Following \citet{Bruna2013}, we apply a classifier based on an affine space model with the approximate affine space determined by Principal Component Analysis (PCA) of each class. To be specific, let $SX_k$ denote the set of MST coefficients for all of our images belonging to class $k$. $SX_k$ can be organized into a $N_{i,k}\times P$ matrix where $N_{i,k}$ is the number of images available for class $k$ and $P$ is the number of scattering coefficients (\textit{i.e.}, the coefficients have been stacked into a vector of length $P$). The columns of $\Delta_k$ may be transformed to have zero mean for each of the $P$ coefficients $\Delta_k = SX_k - \mathbb{E}(SX_k)$. We may then perform principal component analysis on $\Delta_k$ by finding the eigenvectors $\{\mathbf{U}_{j,k}\}_{j=1}^P$ and corresponding eigenvalues $\{\Lambda_j\}_{j=1}^P$ of the covariance matrix $\Delta_k^T\Delta_k$. Taking $\mathbf{U}_{j,k}$ to be ordered such that $\Lambda_j > \Lambda_{j+1}$, we keep only the first $d \ll P$ principal vectors $\{\mathbf{U}_{j,k}\}_{j=1}^d$. Letting $\mathbf{V}_k = \text{span}(\{\mathbf{U}_{j,k}\}_{j=1}^d)$,  we may construct the affine approximation space for class $k$
\begin{equation}
\label{eqn:affinespace}
\mathbf{A}_k = \mathbb{E}(SX_k) + \mathbf{V}_k.
\end{equation}
Finally, for a new image with scattering coefficients $Sx$, the class assigned to the image is given by 
\begin{equation}
\hat{k}(x) = \underset{k}{\operatorname{argmin}} || Sx - P_{\mathbf{A}_k}(Sx)||.
\end{equation}

\begin{figure}[ht]
\center\includegraphics[width=\columnwidth]{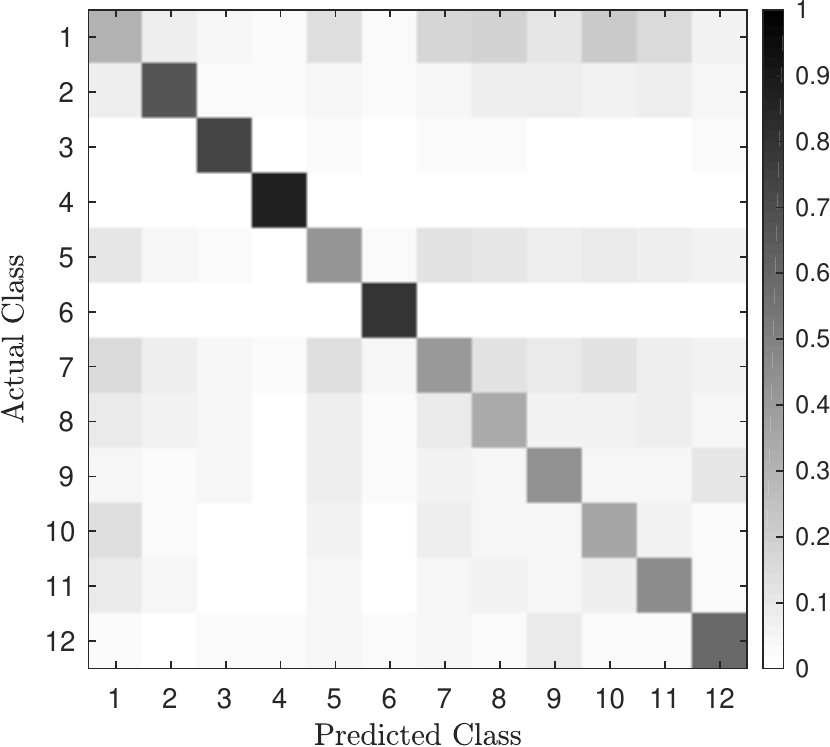}
\caption{\label{fig:confusion} The surrogate confusion matrix $\Omega_{ij}$ defined by Eq.~\eqref{eqn:sepdec} which demonstrates that the constructed double helix classes are well separated in the MST space.}
\end{figure}

In order to evaluate how effectively the classes are separated one may define the ratio of the expected value of the distance of class $i$ to the affine space for class $j$ divided by the expected value of the distance of class $i$ to its own affine space, 
\begin{equation}
\label{eqn:sep}
R^2_{ij} \equiv \frac{E(|| SX_i - P_{\mathbf{A_j}}(SX_i)||^2)}{E(|| SX_i - P_{\mathbf{A_i}}(SX_i)||^2)}.
\end{equation}
Note that if the classes are well separated, then $R_{ij}^2$ will be very large for $i\neq j$, while $R_{ii}^2=1$. It thus makes sense to define the matrix
\begin{equation}
\label{eqn:sepdec}
\Omega_{ij} =N_j e^{-|R_{ij}|},
\end{equation}
where $N_j$ is a column-wise normalization ensuring that each column of $\Omega_{ij}$ sums to $1$.  The off-diagonal elements are indicative of overlap among the the tails of the class distributions.  Conservative assumptions are made that the distribution is exponential, not Gaussian, and the algebraic geometric factor is ignored.  Both lead to estimation of more distribution overlap.  The assumption of an exponential distribution leads to the simplification of the same form for the cumulative distributions.  Equation~\eqref{eqn:sepdec} gives a rough probability that members of class $j$ would have values that would be classified as class $i$, that is the confusion matrix, $P(C_i | C_j)$.  For the case that there is small overlap in the class distribution and there are limited samples, $\Omega_{ij}$ is a high fidelity surrogate for the confusion matrix.  This is because this statistic calculates moments of cluster size and separation, rather than the occurrence of the rare cluster overlap events.  Figure~\ref{fig:confusion} shows the matrix $\Omega_{ij}$ for our case demonstrating good class separation as indicated by the fact that the matrix is strongly diagonal.  The chance of miss-classification is extremely small ($<0.1\%$), and the average class precision is $0.00017$ while the average class separation is $10$.  Here we have used the definitions of class separation, $R_d^2$, and precision, $r_d^2$, given in \citet{Bruna2013},
\begin{equation}
\label{eqn:sep2}
R^2_d \equiv \frac{1}{N_c} \sum_{i=1}^{N_c}  \frac{E(\text{min}_{j\ne i}|| SX_i - P_{\mathbf{A_j}}(SX_i)||^2)}{E(|| SX_i - P_{\mathbf{A_i}}(SX_i)||^2)},
\end{equation}
\begin{equation}
\label{eqn:prec}
r^2_i \equiv \frac{E(|| SX_i - P_{\mathbf{A_i}}(SX_i)||^2)}{E(|| SX_i ||^2)}
\end{equation}
and
\begin{equation}
\label{eqn:prec2}
r^2_d \equiv \frac{1}{N_c} \sum_{i=1}^{N_c}  r^2_i.
\end{equation}
Note that the separation is just the average of the separation matrix given in Eq.~\eqref{eqn:sep}.  The geometric meaning of the separation and precision are shown in Fig.~\ref{fig:classes}.  Also note in Fig.~\ref{fig:classes} that the classes were constructed so that the separation is about 6 to 10 in $\theta_i$-space.  This should be the upper limit on the class separation of any classifier.  The fact that the realized class separation with an affine classifier is the optimal value of 10, indicates that the MST metric has exposed the model parameters as linear combinations of the MST metric and no better representation could be found.

\begin{figure}[ht]
\center\includegraphics[width=\columnwidth]{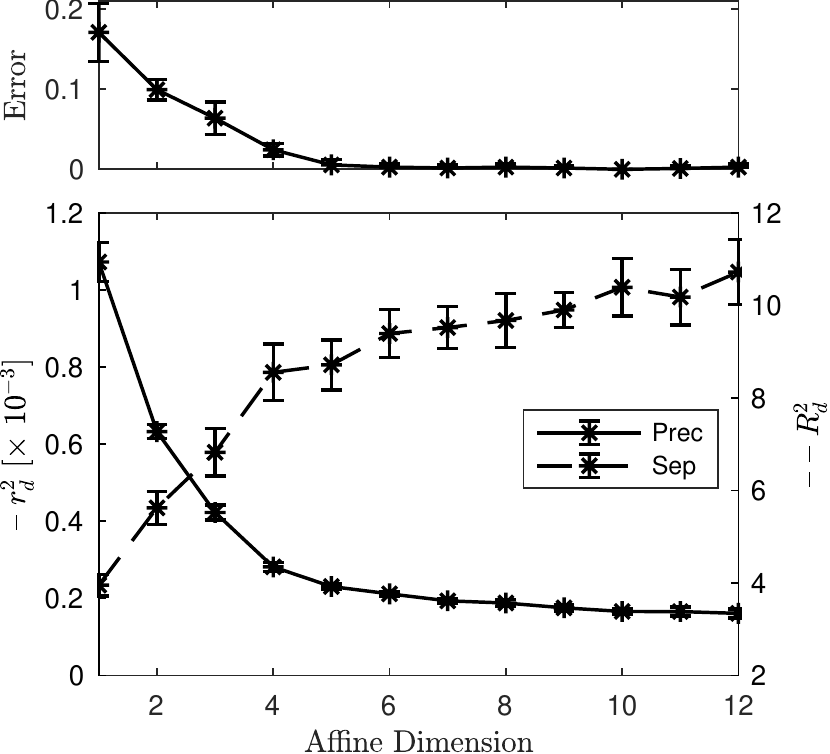}
\caption{\label{fig:affine_opt} Optimization curve for the dimension of the affine space. As the dimension of the affine space increases the error decreases and the average intra-class precision ($r^2_d$) decreases while the inter-class separation ($R^2_d$) increases.}
\end{figure}

Increasing the dimension of the PCA affine approximation space has been shown to make the MST more robust to rotations by effectively reducing the spread of the intra-class affine space while increasing the spread of the inter-class separations for classification problems using an affine classifier and the MST.\citep{Bruna2013} Here we observe a similar effect of affine space dimensionality on performance as demonstrated in Fig.~\ref{fig:affine_opt}. While the performance is maximized for a dimension of 10, there are diminishing returns after a dimension of 4.

\subsection{\label{sec:regression}Regression model and ML pipeline}

\begin{figure*}[ht]
\includegraphics[width=2\columnwidth]{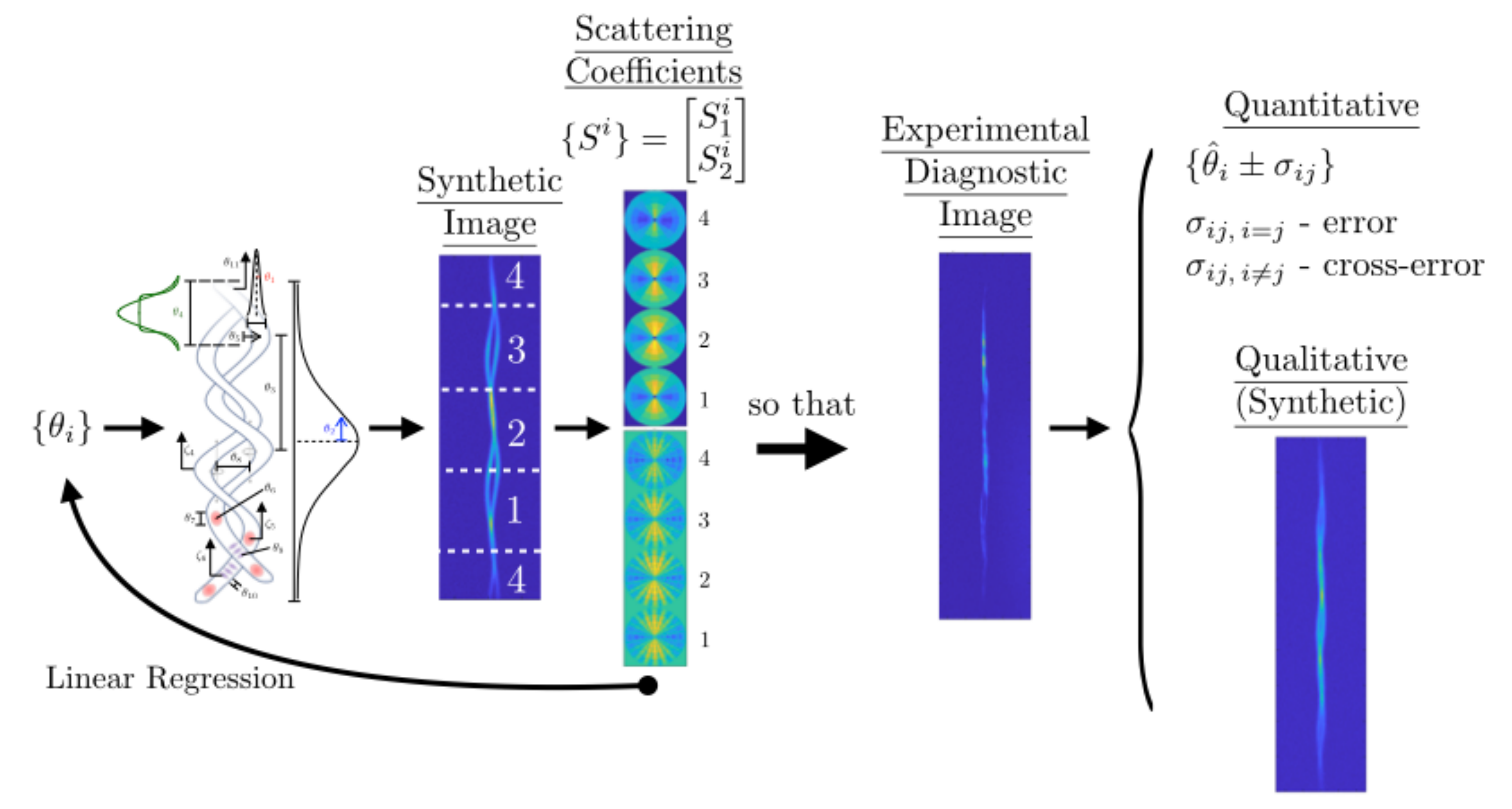}
\caption{\label{fig:mstr} The MST regression pipeline for morphology characterization of experimental stagnation images.  Starting at the left of this figure, an ensemble of synthetic stagnation images are formed from the parameterized model.  The MST is taken of this ensemble over the four numbered patches, then the MST is regressed to give the synthetic model parameters.  This regression is then applied to an experimental diagnostic image to give estimates of the model parameters with uncertainty including correlation.  The synthetic image using the Maximum A posteriori Probability (MAP) parameters can then be constructed.   Ensemble of synthetic stagnation images used for regression (shown via animation in this link to a  \href{https://youtu.be/uqx-ZkV6TxE}{Multimedia View}).  Shown in this animation are the individual members of the ensemble (left), and the corresponding synthetic stagnation image constructed from the MAP parameters regressed from the MST of the member of the ensemble (right).}
\end{figure*}

We now consider the regression problem as highlighted in Fig.~\ref{fig:mstr}.  The goal of this regression is to estimate the synthetic model parameters with uncertainty given an experimental image.  This Machine Learning (ML) pipeline takes as input an image of a plasma stagnation column and outputs a set $\{\theta_i\}_{i=1}^{11}$ characterizing the morphology of the column along with an estimate of the uncertainty of the output. This will be achieved by creating a set of synthetic images from Eq.~\eqref{eqn:h} using a large set of randomly chosen $(\theta_i,\zeta_i)$, computing the MST, and performing a regression from MST coefficients to $\theta_i$.  Note the absence of $\zeta_i$ in our output as those are meant to represent unimportant transformations, such as rotating the viewing angle, which does not alter the fundamental morphology.  Specifically, image realizations are produced, using the synthetic model, from a random sampling of the log-uniformly distributed model parameters. The statistical properties of these distributions are determined by visually confirming that helices produced encompass what is reasonable to expect from experiment. Additionally, most of the quantities we wish to learn from the helical images (\textit{i.e.} the $\theta_i$'s) are non-negative. As a result, we chose to $\log_{10}$ scale all of the $\theta_i$ values except for the strand phase $\theta_{11}$.  The log scaling of the values is equivalent to assuming a log-normal prior distribution.  This choice of using a log-normal distribution is verified in Sec.~\ref{sec:metric_verify}.

Before conducting a linear regression from ($\log_{10}$ scaled) MST coefficients to (scaled) helical parameters, we standard normal scale $\theta$ and $\mathbf{S}$. We will henceforth refer to the transformed quantities as $\tilde \theta$ and $\mathbf{\tilde S}$. Principal Component Analysis (PCA) is then employed to find a set of orthonormal basis vectors by which to rotate the MST coefficients and model parameters into a more directly correlated space, prior to linear regression,  by applying Singular Value Decomposition (SVD) to the cross-covariance between the MST coefficients and model parameters from the training set, CCOV$( \mathbf{\tilde \theta}, \mathbf{\tilde S}) = \mathbf{\tilde \theta}^T\mathbf{\tilde S}/(N-1)$. Here, $N$ is the number of training samples used to construct the cross-covariance. The SVD factors the cross-covariance matrix into a set of transformation matrices $\mathbf{U}$ and $\mathbf{V}$ bounding a diagonal matrix $\mathbf{\Sigma}$ containing a set of singular values
\begin{equation}
\label{eqn:svd}
\mathbf{U \Sigma V}^T = \frac{\mathbf{\tilde \theta}^T\mathbf{\tilde S}}{N-1}.
\end{equation}

\begin{figure*}[ht]
\includegraphics[width=2\columnwidth]{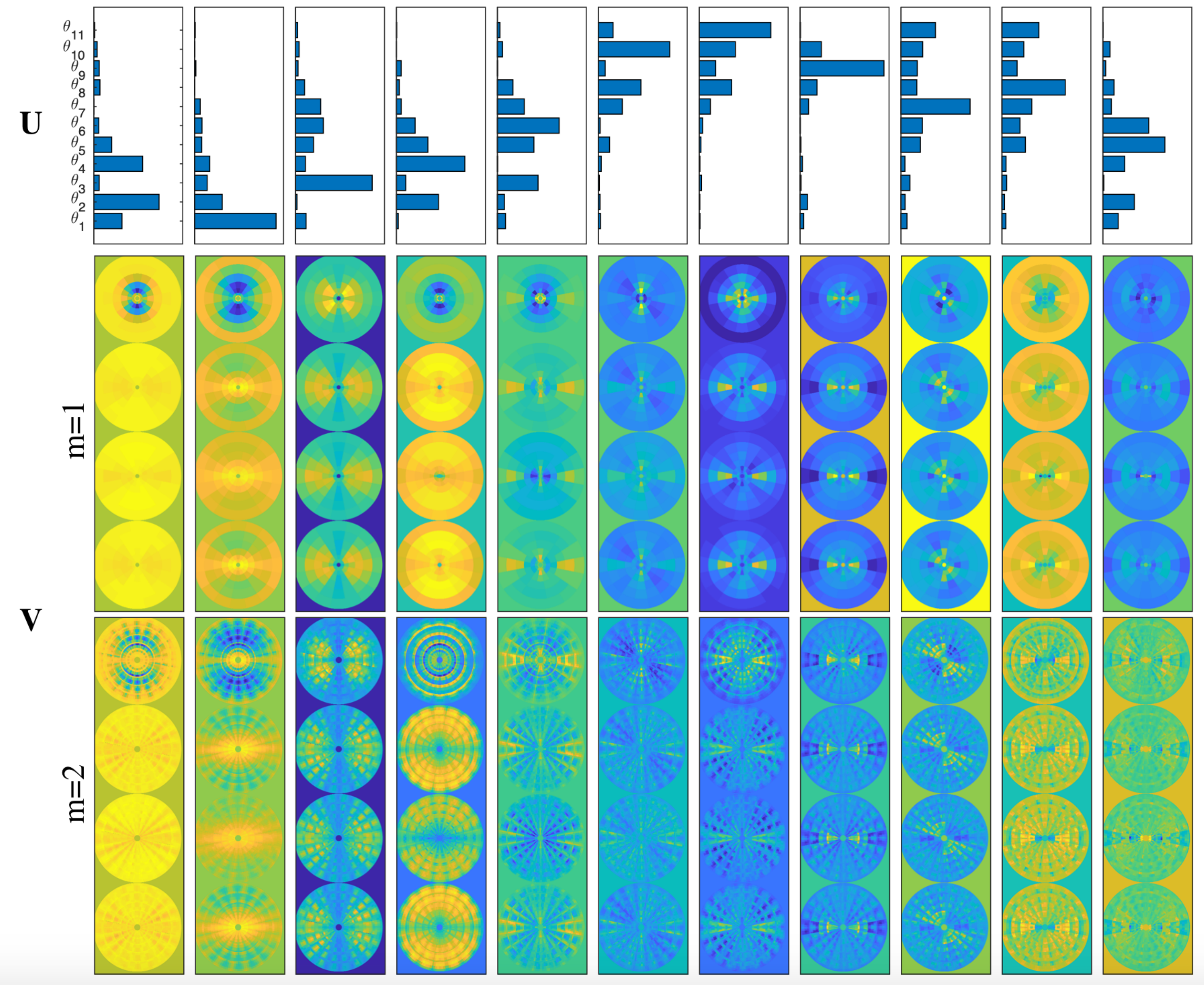}
\caption{\label{fig:pcv} Orthogonal basis vectors, $\mathbf{U}$ (top panel) and $\mathbf{V}$ (bottom panel), which map the model parameters and scattering coefficients, respectively, into the directly correlated space.  For the scattering coefficients, the first order is shown in the top row and the second order in the bottom row.  Each column represents one of the orthogonal basis vectors from most to least correlated (left to right). For some of these vectors there is a simple interpretation such as the second from the left -- $\theta_1$ and a vertical striping in the image.  For others it is a combination of $\theta_i$ and a rich texture.  For these cases further insight to the texture could be given by taking the inverse MST.}
\end{figure*}

\begin{figure}[ht]
\includegraphics[width=\columnwidth]{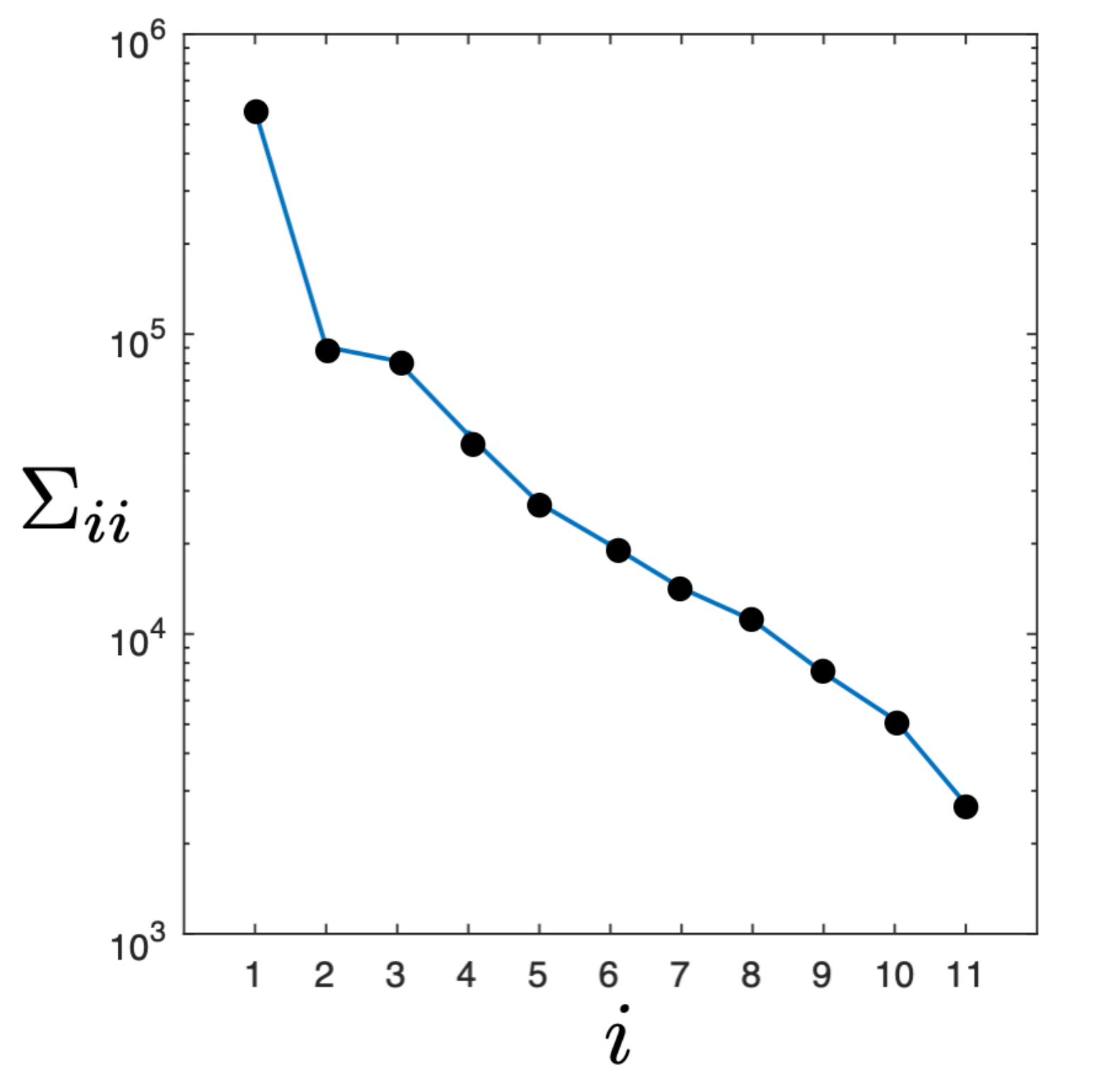}
\caption{\label{fig:sv} Singular values of the cross-covariance matrix, given by the diagonal elements of $\Sigma$.  Gives the significance of the orthogonal vectors of Fig.~\ref{fig:pcv}.  The greater the singular value of the cross-covariance the larger the contribution to the regressor's prediction.}
\end{figure}

PCA is often used on linear systems for dimensionality reduction, however for reasons that will be discussed shortly, we retain full dimensionality. This set of transformation matrices provides a set of orthogonal basis vectors along which $\tilde{\theta}$ and $\tilde{\mathbf{S}}$ are most directly correlated, ordered from strongest to weakest correlation (see Fig.~\ref{fig:pcv} and Fig.~\ref{fig:sv}).  Most of the correlation is contained in the first four dimensions, about 91\% of the variation. The model parameters $\tilde{\theta}$ and scattering coefficients $\tilde{\mathbf{S}}$ are rotated into the directly correlated space such that $\mathbf{Y}=\mathbf{\tilde \theta U}$ and $\mathbf{X}=\mathbf{\tilde S V}$ define the rotated variables.

Regressing the rotated scattering coefficients $\mathbf{X}$ back onto the rotated $\mathbf{Y}$ is accomplished using multidimensional linear regression,  
\begin{equation}
\label{eqn:reg}
Y_j = b_j + \sum_{i=1}^{p}X_i m_{ij} + \epsilon_j,
\end{equation}
where $\mathbf{m}$ is the map from $\mathbf{X}$ to $\mathbf{Y}$ (\textit{e.g.}, ``slope''), $\mathbf{b}$ is the bias (\textit{e.g.}, intercept), $\mathbf{\epsilon}$ is the error term, and $\epsilon =\mathcal{N}(0,\mathbf{\Lambda})$ is assumed to be a zero mean normal random variable with covariance matrix $\mathbf{\Lambda}$. Writing Eq.~\eqref{eqn:reg} in matrix notation, the bias is absorbed into the slope such that $\mathbf{Y}=\mathbf{XM}+\mathbf{\epsilon}$. 

Note that Eq.~\eqref{eqn:reg} implies that the prediction for a new input $\mathbf{X}$ is $\mathbf{Y}_\text{pred} =\hat{\mathbf{Y}} = \mathbf{XM}$ since $\overline{\mathbf{\epsilon}}=0$. Importantly, this would also be able to characterize the uncertainty in our prediction if we had an estimate of $\mathbf{\Lambda}$. In order to estimate $\mathbf{M}$ and $\mathbf{\Lambda}$, note that Eq.~\eqref{eqn:reg} specifies a likelihood function 
\begin{equation}
\begin{split}
P(\{\mathbf{Y}_i\}|\mathbf{M},\mathbf{\Lambda},\{\mathbf{X}_i\}) = \prod_{i=1}^{N} \frac{1}{\sqrt{(2\pi)^k|\mathbf{\Lambda}|}}\\
\times e^{-\frac{(\mathbf{Y}_i-\mathbf{X}_i\mathbf{M})^T\mathbf{\Lambda}^{-1}(\mathbf{Y}_i-\mathbf{X}_i\mathbf{M})}{2}},
\end{split}
\end{equation}
where the training data are assumed independent and identically distributed (i.i.d.) and $k$ is the dimensionality of our output space (here $k=11$ since there are $11$ theta parameters to which we wish to regress).
A maximum likelihood estimate of the coefficients of the matrix $\mathbf{M}$ and error covariance matrix $\mathbf{\Lambda}$ are determined by finding their values which maximize the likelihood function over our training data. Equivalently, since the logarithm is monotonic, we may maximize the log-likelihood $\mathcal{L}$. The solution is derived in many statistics and machine learning textbooks (see \textit{e.g.} \citet{bishop}) and is given by
\begin{equation}
\label{eqn:mleM}
\mathbf{M}_\text{MLE} = (\mathbf{X}^T\mathbf{X})^{-1}\mathbf{X}^T\mathbf{Y},
\end{equation}
which is the typical ordinary least squares solution where $\mathbf{X}_i$($\mathbf{Y}_i$) have been stacked to create $\mathbf{X}$($\mathbf{Y}$) and the error covariance matrix is
\begin{equation}
\label{eqn:mleL}
\mathbf{\Lambda}_\text{MLE}= \frac{1}{N} \sum_{i=1}^N(\mathbf{Y}_i-\mathbf{X}_i\mathbf{M}_\text{MLE})^T(\mathbf{Y}_i-\mathbf{X}_i\mathbf{M}_\text{MLE}),
\end{equation}
which is just the estimate of the population covariance matrix of the difference $(\mathbf{Y}-\mathbf{Y}_\text{pred})$.

For a new image, we can now estimate a set of values $\theta$ along with an estimate of the uncertainty on theta according to the following algorithm.  We start with the MST feature extraction:
\begin{enumerate}
  \item Compute the first and second order scattering coefficients of the image on a 4x4 grid (see Figure~\ref{fig:grid}).
  \item Discard all but the second column from the grid for each of the 2 sets of coefficients.
  \item Compute $\log_{10}$ of the scattering coefficients and flatten into a vector to get $\mathbf{S}$.
  \item Standard normal scale using the mean and standard deviation estimated on the training set to get $\mathbf{\tilde S}$.
  \item Project onto principal components to get $\mathbf{X} = \mathbf{\tilde S} \mathbf{V}$.
  \end{enumerate}
  We then do the regression:
  \begin{enumerate}
  \item Compute $\mathbf{Y}_\text{pred} = \mathbf{X}\mathbf{M}_\text{MLE}$.
  \item Create a set of values consistent to within the error term 
  \begin{equation*}
  \{\mathbf{Y}_\text{pred,i}\}_{i=1}^{N_\text{resamp}} = \mathbf{Y}_\text{pred} + \{\mathcal{N}_i(0,\mathbf{\Lambda}_\text{MLE})\}_{i=1}^{N_\text{resamp}}.
  \end{equation*}
  \item Compute $\{\tilde \theta_i\} = \{\mathbf{Y}_{\text{pred},i}\mathbf{U}^{-1}\}$.
  \item Compute $\{\theta_i\}$ by inverting standard normal scaling of $\tilde \theta$ using the mean and standard deviations of $\tilde \theta$ computed from the training set and then invert the $\log_{10}$ scaling performed on all but the last component of $\theta$.
  \item We now have an estimate of the distribution of $\theta$ consistent with the original image. We may report the prediction and error as means and standard deviations, or as percentiles (\textit{e.g.}, report the $50^{th}$ percentile as the prediction and the $2.5$-percentile and $97.5$-percentile as lower and upper bounds).  The estimates of the distribution are subject to the caveats expressed at the end of this section.
\end{enumerate}
In our case, inverting the transformations leads to an asymmetric distribution of $\theta$ values consistent with the original image, so here we will report the $95\%$ confidence interval and the mode of the distribution rather than mean and standard deviation for any predictions.

Before moving on to discuss results, we note that the cross-covariance matrix computes the set of basis vectors along which the quantities $\tilde{\theta}$ and $\tilde{\mathbf{S}}$ exhibit the strongest linear correlation.  As a result, any nonlinear relationships between $\tilde{\theta}$ and $\tilde{\mathbf{S}}$ will not be recoverable upon linear regression.  First attempts using a linear regression given by Eq.~\eqref{eqn:mleM}, when truncating the dimensionality of the principle components to the number of singular values, showed nonlinear bows.  A more generalized model was constructed to capture this nonlinear behavior by including the full SVD.  The nonlinear aspects are captured by the nonlinear dependence of the additional SVD components on the reduced set of SVD components.

A slight modification to the predictive model is required to mitigate numerical issues with this more generalized model. This is a repercussion of using the full SVD on the cross-covariance matrix which causes the quantity $\mathbf{X}^T\mathbf{X}$ from Eq.~\eqref{eqn:mleM} to be ill-conditioned. Applying $L2$-regularization to the predictive model is shown to be an effective mitigation procedure. The predictive model with $L2$-regularization is
\begin{equation}
\label{eqn:predictive_model}
\hat{\mathbf{Y}} = \mathbf{X} (\mathbf{X}^T\mathbf{X} + \lambda \mathbf{I})^{-1}(\mathbf{X}^T\mathbf{Y})
\end{equation}
where $\lambda$ is optimized through cross-validation, maximizing $R^2$, where 
\begin{equation}
R^2 \equiv 1 - \frac{\sum_i (\mathbf{Y}_i - \hat{\mathbf{Y}}_i)^2}{\sum_i (\mathbf{Y}_i - \bar{\mathbf{Y}})^2}.
\end{equation}
We find that the optimum value of $\lambda_{CV}$ is $0.005614$.  Note that the regularization is an assumption on the gradients or correlation.  This does a have an effect on the estimate of the uncertainty as will be discussed at the end of this subsection.

Image realizations ($N=2048$) are produced, using the synthetic model, from a random sampling of the log-uniformly distributed model parameters.  The statistical properties of these distributions are determined by visually confirming that helices produced encompass what is reasonable to expect from experiment.  We chose the log-uniform distribution over a uniform distribution, because we were not certain of the order of magnitude of the model parameters.  If we would have chosen a uniform distribution, we would have not only under sampled the smaller scales, we would have biased the estimation to the largest scale in the population.  For each $\mathbf{\theta}$ realization, a set of features is extracted from its corresponding synthetically generated image using the MST. The data set is randomly separated into a training ($50\%$), validation ($25\%$), and test ($25\%$) sets. The training set is used for model training, the validation set is used for cross-validation and model selection, while the test set aside and used to asses the performance of the selected model.

\begin{figure*}[ht]
\includegraphics[width=2\columnwidth]{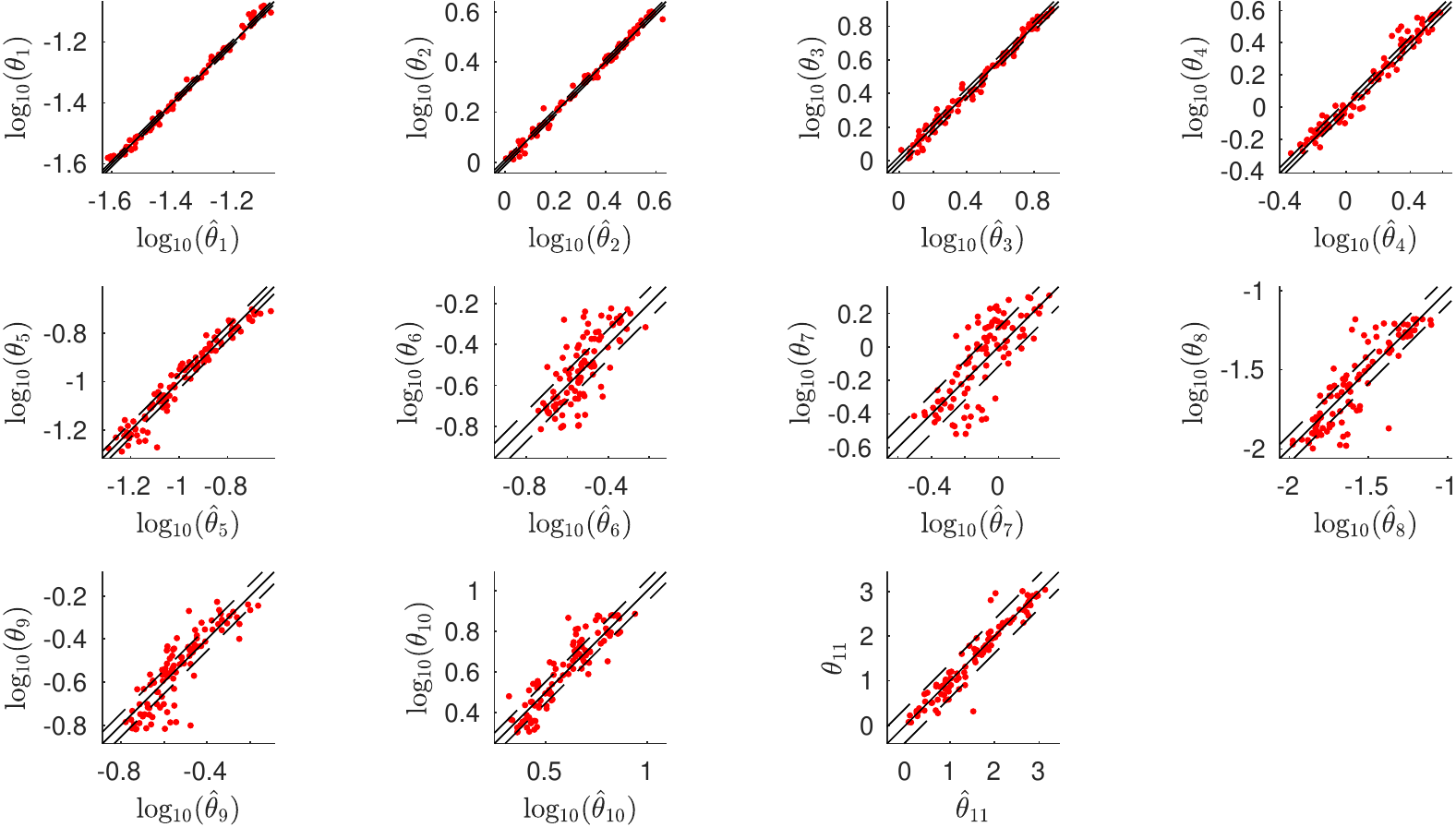}
\caption{\label{fig:mstr_performance_param} MST regressor performance. The scatter plots show predicted vs. actual parameters in $\log_{10}$-space. The first several demonstrate very good performance, while the later components show slightly less performance. This may be indicative of nonlinearity which the linear regression cannot explain, or it may be variance caused by the unexplained $\zeta$ parameters.  Units for all $\theta_i$ are arbitrary, but consistent with those shown in the histograms of Fig.~\ref{fig:classes}.}
\end{figure*}

\begin{figure}[ht]
\includegraphics[width=\columnwidth]{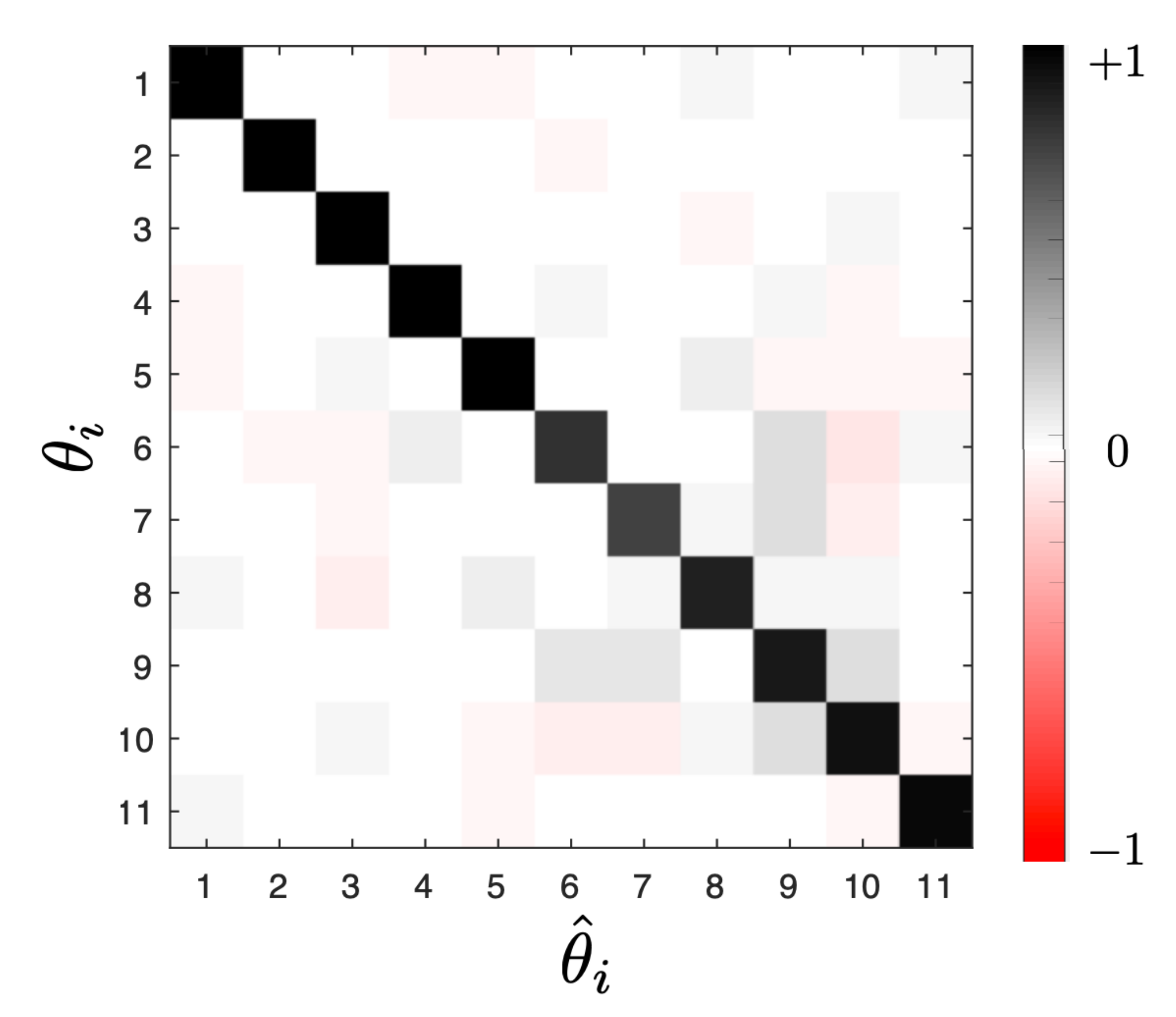}
\caption{\label{fig:mstr_performance_param_2} Further look at the MST regressor performance. Correlation plot which shows that the correlation of the predicted parameters to the actual parameters is very diagonal and close to 1.}
\end{figure}

The scatter plots in Fig.~\ref{fig:mstr_performance_param} and Fig.~\ref{fig:mstr_performance_param_2} show predicted vs. actual morphological parameters of the test set in the $\log_{10}$-scaled MST coefficient space.  There is reasonable agreement over a large range of parameter space.  The correlation is very diagonal and close to 1, and the regression coefficient is quite good, $R^2 = 0.91$.  This all shows that the regression is performing very well.

Before moving on, we need to discuss the uncertainty that is estimated by this regression.  The uncertainty in the regression due to uncorrelated experimental noise is captured (if the experimental noise is Gaussian and of the size that was modeled), but not due to correlated experimental noise.  While estimation uncertainty due to the stochastic nature of synthetic images is captured, both random and systematic error due to the limited nature of these synthetic images compared to the experimental images and due to incorrect statistical assumptions in this analysis are not.  Because of this, the random error could be larger than estimated and there could be systematic error or bias in the estimate.  Interestingly this leads to an iterative approach to improving the models to reduce the systematic error and correct for the underestimation of the random error.  When the results of this regression are applied to experimental or simulation images and compared to a physical theory, a systematic bias and/or greater scatter in the regressed model parameters than predicted by the undertainty could be observed.  This is a symptom that the regression model and/or the physical model should be improved.  Iterative improvements to the models can be made to remove these pathologies in the error.

\subsection{\label{sec:metric_verify}Metric design verification}
Table \ref{tab:optimization} shows cross validation results which aided in the design of the MST metric. The \textit{base} metric was constructed using the aforementioned design criteria, most of which were inherited from previous image classification applications of the MST.  We found a modest improvement in performance when using a much larger training set (four times bigger).  From our cross validation, there was a modest drop in performance when not $\log_{10}$-scaling the MST coefficients, not using the second order MST ($m=1$ only), using integrated intensity instead of max value normalization, and decreasing the number of MST filter rotations.

\begin{table}
\caption{\label{tab:optimization}Cross-Validation results. All models are slight deviations from the Base Model defined by 2048 training images, max normalization, $m$=2, $4 \times 4$-gridding, 8 rotations and $\log_{10}$-features.}
\begin{tabular}{lcc}
\\
\textbf{Model} && \textbf{Validation Set} $\mathbf{R}^2$\\
\hline
Base && $0.9094$\\
\hline
$8192$ training images && $0.9237$\\
$m=1$ && $0.7238$\\
4 rotations && $0.8449$\\
non-$\log_{10}$ on features && $0.8752$\\
integrated intensity normalization && $0.8849$\\
\hline
\end{tabular}
\end{table}

\section{\label{sec:results}Applications and results}
There are two primary cases of interest for applying our method. The first is to be able to quantitatively compare experimental data to simulation. The second is to be able to compare morphology between different experiments and quantify what those differences are. In doing so, we will be able to make statistically sound inferences about discrepancies in morphology. By providing this capability, the method will provide physical insight into the physical mechanisms causing the differences. To this end, we conduct some initial studies which show how the method will be used.

\subsection{\label{sec:sim_exp}Simulation-to-experiment comparison}
\begin{figure}[ht]
\includegraphics[width=\columnwidth]{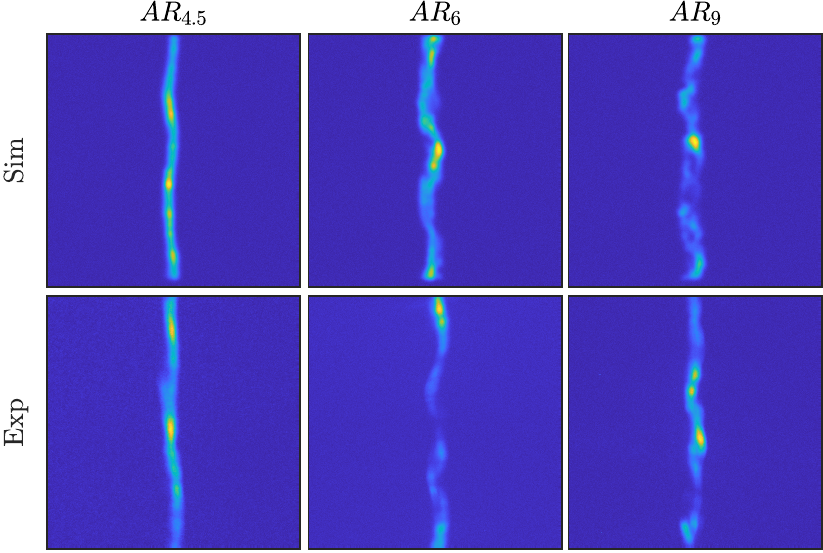}
\caption{\label{fig:sim_exp_stag} Stagnation images from selected MagLIF experiments at varying aspect ratios and their corresponding simulated (\texttt{GORGON}/MHD) counterparts.}
\end{figure}

The experimental images are obtained from the Continuum X-ray Imager instrument fielded on Z.\citep{Gomez2014}  We include self-emission images from AR4.5, AR6 and AR9 MagLIF experiments fielded on Sandia's Z-Machine -- experiments z3017, z2839 and z3018, respectively.\citep{Ampleford2019}  For each of the experimental images, synthetic x-ray self-emission images are taken from 3D radiation magnetohydrodynamic (rad-MHD) \texttt{GORGON} \citep{Chittenden2004} simulations modeling a corresponding experiment. These simulations are run with continuous virtual boundary edges at a height of $5$ $mm$ and the synthetic images are calculated using a ray-tracing algorithm onto a virtual image plate. The experimental images have been vertically cropped down to a $5$ $mm$ height to compare with their simulated counterparts.  The experimental images continue to have similar structure over their full height of about $1$ $cm$.  Figure \ref{fig:sim_exp_stag} shows the comparison of simulated and experimental self-emission images at several different liner aspect ratios.

There are distortions to both the experimental and simulation x-ray self-emission images.  For the experimental images there are instrumental responses, noises, and calibrations that can not be explicitly estimated.  For the simulations there are approximations to the physics, and numerical error in the calculations.  This leads to the ``true'' image being shifted into different \textit{domains} for the experiments and the simulations. In order to address this discrepancy, we have developed a \textit{background subtraction} method. The method works by projecting out a background vector $\mathbf{B}_1$ given by the first principal component of the covariance between simulation and experiment, such that $\tilde{S}^\text{AR}_\text{domain} = S^\text{AR}_\text{domain} - \text{proj}(S^\text{AR}_\text{domain}, \mathbf{B}_1)$, where AR is the aspect ratio, and the domain represents whether the features are from simulated or experimental images.  Physically, this assumes that the dominate difference between an experiment and its corresponding simulation is due to this experimental and/or simulation distortion.  This method will be verified by an increase in the class separation and better precision after the background subtraction is done.

\begin{figure}[ht]
\includegraphics[width=\columnwidth]{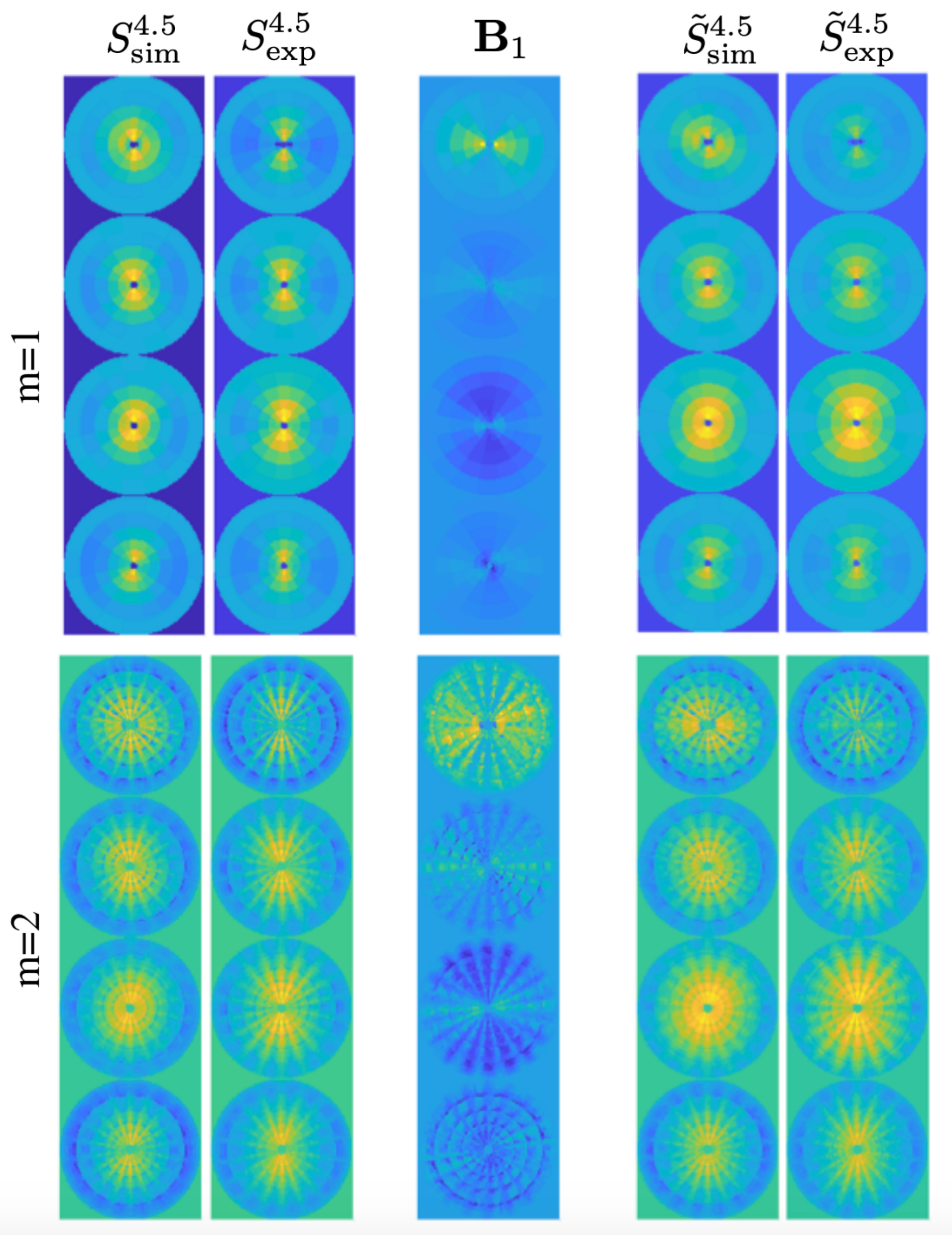}
\caption{\label{fig:bs} Background Subtraction. The top row are the first order MST coefficients, and the second row are the second order MST coefficients.  The two columns on the left are before the background is projected out, the center column is the background derived from the first principal component of the covariance between simulation and experiment, and the right two columns are after the background is projected out.  This is for the AR$4.5$ case.}
\end{figure}

This process of background subtraction is demonstrated in Fig.~\ref{fig:bs}.  The MST coefficients for the AR$4.5$ case are shown for the simulated and experimental data in the left two columns of the figure. There are apparent qualitative similarities of the MST coefficients between the two cases. However, there looks to be some nontrivial background present in the experimental data, which our approach projects out. Specifically, if we take the first principal component of the covariance between simulation and experiment, we find the center column of Fig.~\ref{fig:bs}  -- the background, $\mathbf{B}_1$. After projecting out this background component from the experimental data (the right two columns of Fig.~\ref{fig:bs}), we can observe similarities and differences between the simulation and experimental morphologies by comparing the overall separation of the scattering coefficients, $R_{kl}^2$, computed as pairwise Euclidean distances, $\sigma_{kl}$, normalized by the average intra AR class distance,
\begin{equation}
R_{kl}^2 = \frac{\sigma_{kl}^2}{\text{mean}(\sigma_{ii}^2)}.
\end{equation}
where $k$ and $l$ refer to the AR index of simulation and experiment, respectively.  We can then visualize how well separated they are by plotting
\begin{equation}
\label{eqn:prob_classification}
\Omega_{kl} = N_l e^{-|R_{kl}|},
\end{equation}
which is analogous to the surrogate confusion matrix, $\Omega_{ij}$, defined in Eq.~\eqref{eqn:sepdec}.  Precision and separation can be similarly be defined.  The effectiveness of the background subtraction is quantified by an improvement in the precision from 0.40 to 0.08, and an increase in the separation from 1.9 to 4.8.  There is essentially no cluster separation before the background subtraction is done, and a reasonable separation after it is done.  The character of the background, $\mathbf{B}_1$, in Fig.~\ref{fig:bs} indicates a large wavelength vertical striping.  This is more likely to be a distortion of the experimental image, than a deficiency in the physics of the simulation.

\begin{figure}[ht]
\includegraphics[width=\columnwidth]{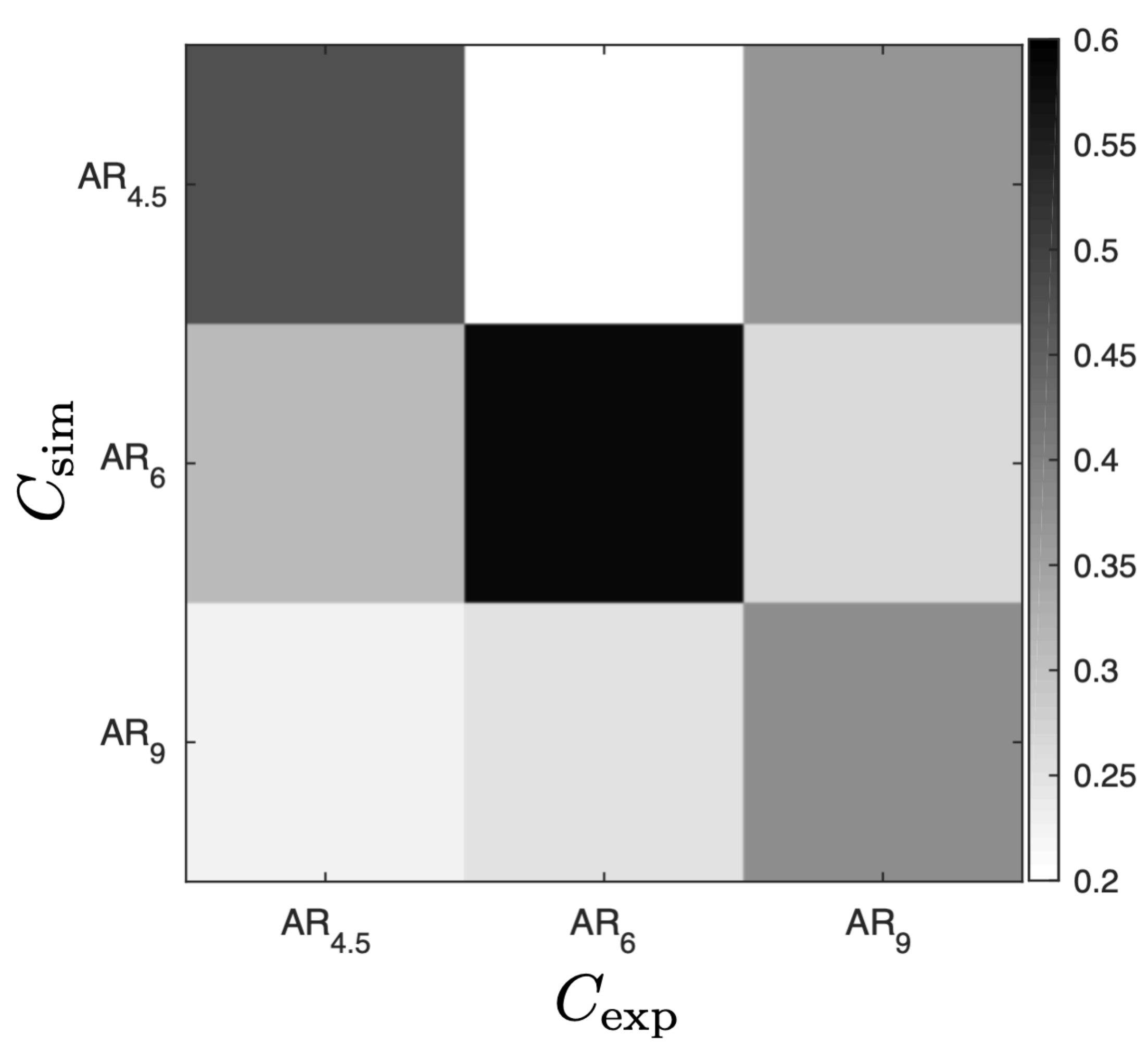}
\caption{\label{fig:sim_vs_exp} The surrogate confusion matrix $\Omega_{kl}$ defined by Eq.~\eqref{eqn:prob_classification} quantifying similarities and differences between the simulations (``sim'') and experiments (``exp'').  Probability of classifying as $C_\text{sim}$ given that it is $C_\text{exp}$, $P(C_\text{sim} | C_\text{exp})$, is plotted as a the image.}
\end{figure}

The quantification of the similarities and differences between the simulations and experiments is shown in Fig.~\ref{fig:sim_vs_exp}.  This demonstrates that the simulations are generally close in MST space to the corresponding experiment, with AR4.5 simulation showing some pairwise similarity to the AR9 experiment.

\subsection{Experiment-to-experiment comparison and analysis}
\begin{figure*}[ht]
\includegraphics[width=2\columnwidth]{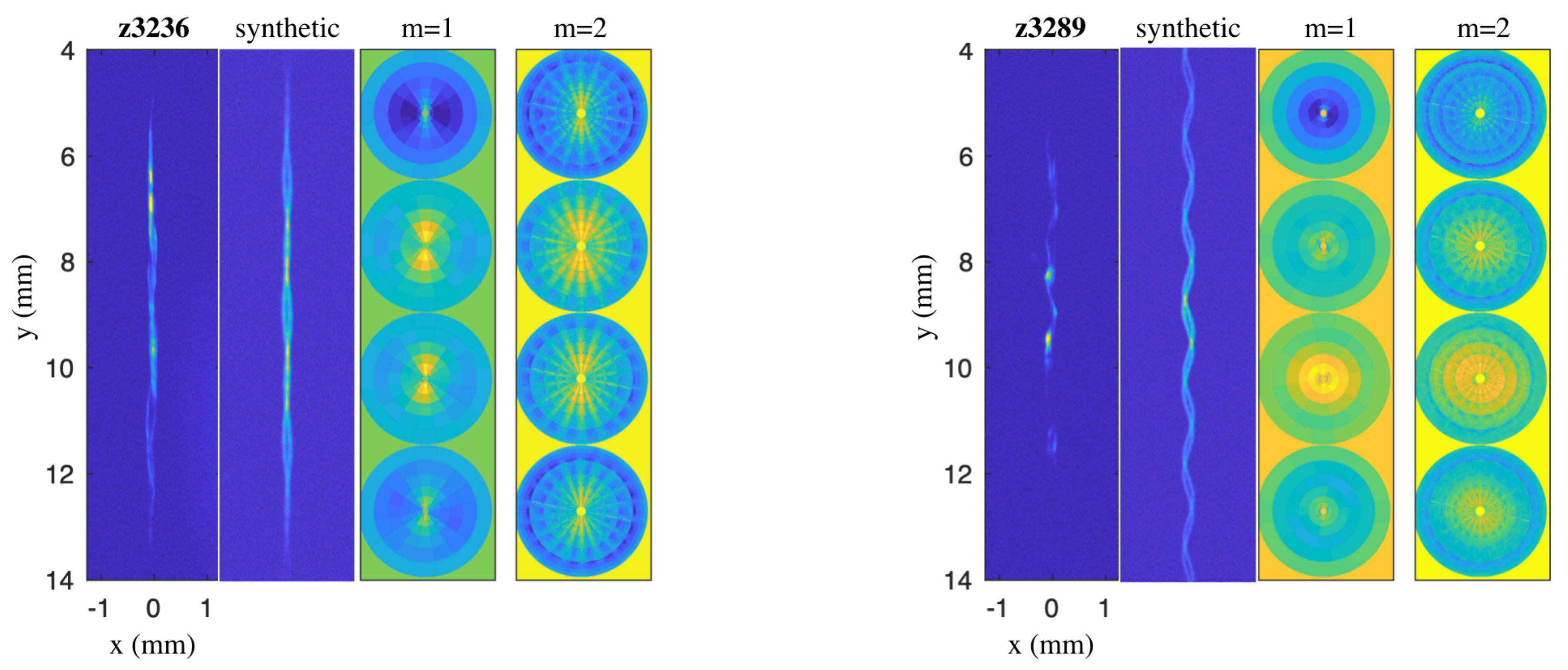}
\caption{\label{fig:coat_vs_uncoat_mst} Comparison of two experiments using the MST.  To the left is shot z3236 with a coated AR9 liner.  To the right is shot z3289 with an uncoated AR6 liner.  Shown, for both cases, are the original stagnation image on the left, the MAP synthetic image, and the MST on the right (both first and second order coefficients).  Ensemble of fit synthetics (shown via animation in this link to a \href{https://youtu.be/e5Dv1VcrOzc}{Multimedia View} for z3236, and a  \href{https://youtu.be/UqxQYHLj6BQ}{Multimedia View} for z3289).  Shown in these animations are the experimental image (left), and member of ensemble of fit synthetics (right).}
\end{figure*}

\begin{figure*}[ht]
\includegraphics[width=2\columnwidth]{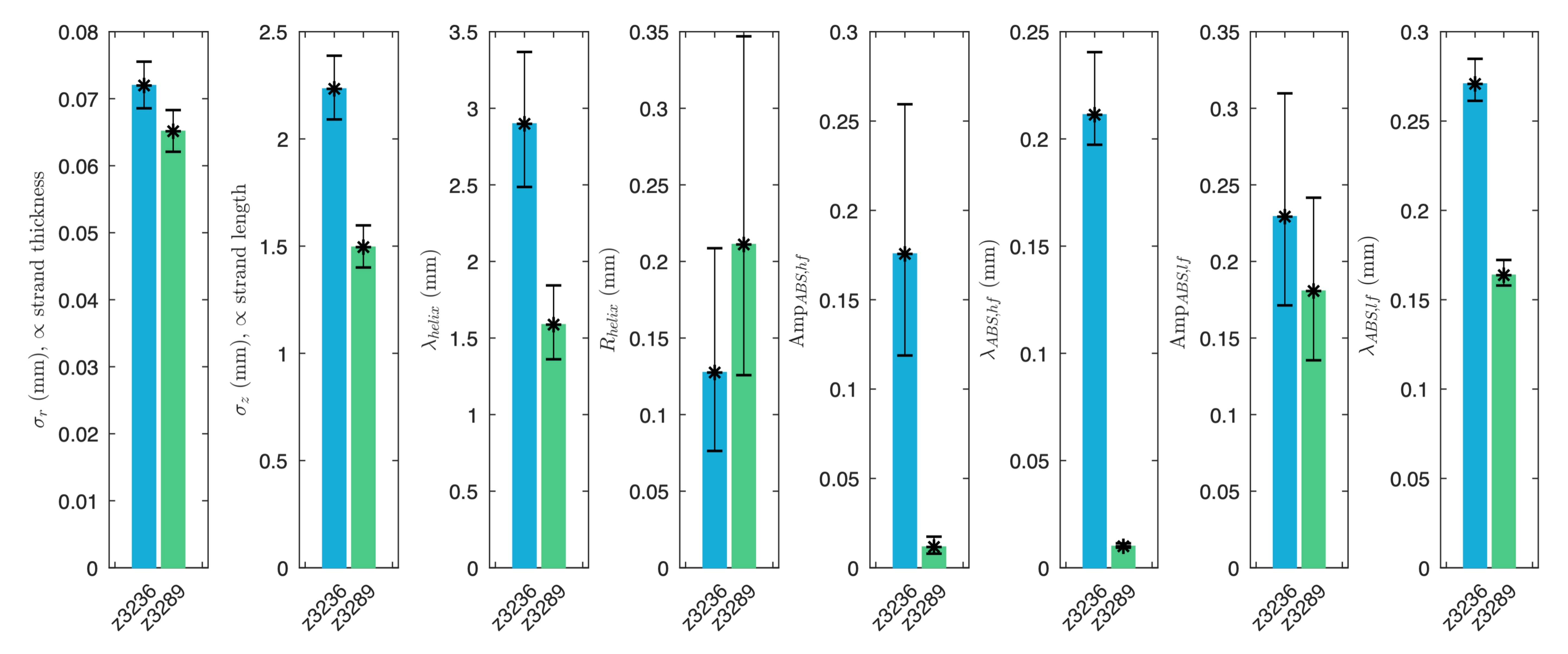}
\caption{\label{fig:coat_vs_uncoat_fit_select} Regressed parameters of the synthetic helical model for the two experiments shown in Fig.~\ref{fig:coat_vs_uncoat_mst}.  The values for shot z3236 are shown in blue on the left, and for shot z3289 in green on the right. Plotted are modes with error bars showing the 95\% confidence interval.  The error bars are asymmetric because this is not in log space. The parameters are (from left to right): strand thickness (mm), strand length (mm), helical wavelength (mm), radius of the helix (mm), amplitude of the high frequency axial brightness perturbations (arbitrary units), wavelength of the high frequency axial brightness perturbations (mm), amplitude of the low frequency axial brightness perturbations (arbitrary units), and wavelength of the low frequency axial brightness perturbations (mm).}
\end{figure*}

Finally, we finish with a discussion of differentiating morphology between experiments. Figure \ref{fig:coat_vs_uncoat_mst} shows the experimental plasma stagnation columns along side the synthetic model for the mean prediction and their first and second order MST coefficients of two different liner designs. To the left is experiment z3236 which utilized a dielectric coated AR9 target, while to the right is experiment z3289 which had an uncoated AR6 liner. The dielectric coating on the exterior of the liner is expected to reduce the amount of magnetic Rayleigh-Taylor growth by reducing the electro-thermal instability that is seeding it.  There are other significant differences between these two experiments in the amplitude of the current drive, preheating laser pulse profile, applied axial magnetic field, magnetic field axial uniformity, and liner configurations. There are obvious differences between the MST coefficients for the two cases; but what is different?  To answer this question, we applied the regression derived in Sec.~\ref{sec:regression}. The regressed synthetic parameters and their uncertainties for MagLIF experiments z3236 and z3289 are shown in Table \ref{tab:coat_vs_uncoat_fit}. The listed uncertainties represent the $95\%$ confidence intervals and are obtained from the multivariate Gaussian distribution of the test data, $\{\mathbf{Y}_\text{pred,i}\}_{i=1}^{N_\text{resamp}}$. The estimates of selected parameters of the synthetic helical model, the $\theta$'s, along with their uncertainties, are plotted for the two cases side-by side in Fig. \ref{fig:coat_vs_uncoat_fit_select}.  For parameters such as radius of the helix and the amplitude of the low frequency axial brightness perturbations, there are negligible differences.  For other parameters such as the strand thickness, there are modest differences.  For yet other parameters such as strand length, helical wavelength, amplitude of the high frequency axial brightness perturbations, the wavelength of the high frequency axial brightness perturbations, and the wavelength of the low frequency axial brightness perturbations; there are significant differences. The reader will also note that the synthetic images given by the mean prediction capture a number of physical features such as the helical wavelength reasonably well.

\begin{table}
\caption{\label{tab:coat_vs_uncoat_fit}MST Regressor determined stagnation column morphological parameters from MagLIF experiments z3236 and z3289.  Units for all $\theta_i$ are arbitrary, but consistent with those shown in the histograms of Fig.~\ref{fig:classes} and in the cross plots of Fig.~\ref{fig:mstr_performance_param}.}
\begin{tabular}{lcccc}
\\
 && \textbf{z3236} && \textbf{z3289}\\
&& \textbf{Coated AR9} && \textbf{Uncoated AR6}\\
\hline
$\theta_{1}$ && $0.0720$ + $(-0.0034, 0.0036)$ && $0.0651$ + $(-0.0031, 0.0032)$ \\
$\theta_{2}$ && $2.2335$ + $(-0.1425, 0.1546)$ && $1.4954$ + $(-0.0950, 0.1013)$ \\
$\theta_{3}$ && $2.1677$ + $(-0.3083, 0.3509)$ && $3.9584$ + $(-0.5638, 0.6389)$ \\
$\theta_{4}$ && $1.5327$ + $(-0.2560, 0.3003)$ && $0.3120$ + $(-0.0509, 0.0622)$ \\
$\theta_{5}$ && $0.0988$ + $(-0.0145, 0.0176)$ && $0.1975$ + $(-0.0282, 0.0337)$ \\
$\theta_{6}$ && $0.2293$ + $(-0.0579, 0.0805)$ && $0.1807$ + $(-0.0452, 0.0609)$ \\
$\theta_{7}$ && $5.3522$ + $(-1.3159, 1.7144)$ && $4.8469$ + $(-1.1991, 1.5604)$ \\
$\theta_{8}$ && $0.0286$ + $(-0.0107, 0.0175)$ && $0.0136$ + $(-0.0051, 0.0084)$ \\
$\theta_{9}$ && $0.1756$ + $(-0.0568, 0.0838)$ && $0.0116$ + $(-0.0037, 0.0058)$ \\
$\theta_{10}$ && $6.8597$ + $(-3.1944, 5.8555)$ && $79.247$ + $(-36.568, 68.318)$ \\
$\theta_{11}$ && $0.4819$ + $(-0.4050, 0.4027)$ && $2.0195$ + $(-0.4009, 0.4079)$ \\
\hline
\end{tabular}
\end{table}

\section{Conclusions}
We have designed and optimized a metric of stagnation morphology using the MST.  This was based on both classification of ensembles of synthetic stagnation images, and regression of those synthetic stagnation images to the morphology parameters used to generate them.  Excellent performance of both the classifier and regressor was obtained.  We demonstrated that the MST provides a convenient basis in which to project out discrepancies between simulated and experimental images.  

This metric is then able to be used to test hypotheses, such as if the AR of the liner makes significant changes to the stagnation morphology, and whether the rad-MHD computer simulations predict the changes to the stagnation morphology.  For the experimental and simulation data analyzed in Sec.~\ref{sec:sim_exp}, the cluster separation was almost 5 (that is, at a 5 sigma level).  The experimental and simulation images in Fig.~\ref{fig:sim_exp_stag} show very subtle differences that could be challenging to quantify without the use of the MST metric, yet show a very significant separation using the metric.  We leave conclusions about the significance of systematic changes in the morphology with AR and the use of dielectic coatings and their prediction by simulations for future publications which will analyze larger and more complete datasets.  

Finally, the regression enabled the morphology parameters of the stagnation to be estimated with uncertainty.  It should be noted that nonlinear aspects of this regression were captured by including more components of the MST SVD vectors in the linear regression of the MST to the morphological parameters.  Before the development of this methodology, only a very rough point estimate of some, not all, of the parameters of the stagnations images could be made (i.e., helical wavelength and strand thickness).  This was complicated by the stochastic nature of the image.  There was no estimate of the uncertainty.  Now, it can be concluded that shot z3236 when compared to shot z3289 has a modest increase in strand thickness, and significant increases in strand length, helical wavelength and the wavelength of the low frequency axial brightness perturbations.  This shot has longer and larger physical structures.  Unfortunately, causal relationships can not be inferred (for instance the effect of the dielectric coating) because of the number of variables that were changed between the shots.  Work is underway to analyze and acquire a larger set of shots with more systematic variation in order to establish the causal relationships.

The MST metric space does look to be a low dimensional representation of the stagnation images.  The affine classifier showed little improvement after about 4 dimensions, and the SVD of the cross variance of the morphology parameters of the synthetic model and the MST contained most of the variance within the first four components.

There are several ways that this research can be improved and expanded upon.  The model has been trained on an inherently 2D synthetic data set whereas the experimental images are projections of complex 3D physical systems onto 2D image plates.  The synthetic model also has a significant amount of symmetry that is not seen in the experimental images. The experimental noise model does not have correlation that might exist in the data. This work could be expanded to use a 3D synthetic model with less symmetry, and use a more realistic noise model based on the experimental data, to address these issues.

Although the use of more components of the MST SVD vectors did reduce the nonlinear artifacts of the linear regression, one could apply some of the modern nonlinear regressions, both shallow and deep.  There are still some experimental images which are too noisy, or exhibit other artifacts which preclude our ability to get reliable morphology parameter estimates.  Additional machine learning and data augmentation methods as well as using a larger data base of experimental images could address this issue.

The connection of the MST to the underlying physics of the rad-MHD, and its emergent behavior is being explored by our ongoing research.\citep{glinsky.et.al.19, glinsky.11}  For example, we are addressing questions such as: what is the formal connection of the MST to physical dynamics of both classical and quantum field systems?  Can a surrogate for the rad-MHD evolution be constructed using the MST, and can the fixed point, that is, emergent behavior be extracted from that surrogate?  What is the expression of helicity and energy in the MST space?  What is the relationship of the MST to algebraic topology, that is concepts such as manifold curvature and the Atiyah-Singer index theorem?\citep{atiyah1963index}

Finally, we emphasize that the \textit{background subtraction} in the MST metric space is essential to obtain a quantitative metric which can be used to compare morphology of simulation and experiment. By studying the variation between and within datasets in this metric space, the distortions and system responses can be characterized and removed.  The simple case that we presented in Sec.~\ref{sec:sim_exp}, using only six samples, demonstrated its potential, but extensions of this work to much larger datasets in future work will provide further clarification on the usefulness of the background subtraction procedure.  It is also not clear what is being compensated by the background subtraction -- distortions to the experimental data, deficiencies in the physics of the simulations, or both.  The character of the background, vertical stripes, is suggestive of an experimental distortion, but there is not enough data to draw any conclusions.

\section{Acknowledgments}
We would like to thank St\'ephane Mallat for many useful discussions, suggestions, and providing a computer software implementation of his Scattering Transformation.  This research was funded by the Sandia National Laboratories' Laboratory Directed Research and Development (LDRD) program.  Sandia National Laboratories is a multimission laboratory managed and operated by National Technology and Engineering Solutions of Sandia LLC (NTESS), a wholly owned subsidiary of Honeywell International Inc., for the U.S. Department of Energy's National Nuclear Security Administration (NNSA) under contract DE-NA0003525.  This paper describes objective technical results and analysis. Any subjective views or opinions that might be expressed in the paper do not necessarily represent the views of the U.S. Department of Energy or the United States Government.  The data that support the findings of this study are available from the corresponding author upon reasonable request.


\begin{thebibliography}{26}%
\makeatletter
\providecommand \@ifxundefined [1]{%
 \@ifx{#1\undefined}
}%
\providecommand \@ifnum [1]{%
 \ifnum #1\expandafter \@firstoftwo
 \else \expandafter \@secondoftwo
 \fi
}%
\providecommand \@ifx [1]{%
 \ifx #1\expandafter \@firstoftwo
 \else \expandafter \@secondoftwo
 \fi
}%
\providecommand \natexlab [1]{#1}%
\providecommand \enquote  [1]{``#1''}%
\providecommand \bibnamefont  [1]{#1}%
\providecommand \bibfnamefont [1]{#1}%
\providecommand \citenamefont [1]{#1}%
\providecommand \href@noop [0]{\@secondoftwo}%
\providecommand \href [0]{\begingroup \@sanitize@url \@href}%
\providecommand \@href[1]{\@@startlink{#1}\@@href}%
\providecommand \@@href[1]{\endgroup#1\@@endlink}%
\providecommand \@sanitize@url [0]{\catcode `\\12\catcode `\$12\catcode
  `\&12\catcode `\#12\catcode `\^12\catcode `\_12\catcode `\%12\relax}%
\providecommand \@@startlink[1]{}%
\providecommand \@@endlink[0]{}%
\providecommand \url  [0]{\begingroup\@sanitize@url \@url }%
\providecommand \@url [1]{\endgroup\@href {#1}{\urlprefix }}%
\providecommand \urlprefix  [0]{URL }%
\providecommand \Eprint [0]{\href }%
\providecommand \doibase [0]{http://dx.doi.org/}%
\providecommand \selectlanguage [0]{\@gobble}%
\providecommand \bibinfo  [0]{\@secondoftwo}%
\providecommand \bibfield  [0]{\@secondoftwo}%
\providecommand \translation [1]{[#1]}%
\providecommand \BibitemOpen [0]{}%
\providecommand \bibitemStop [0]{}%
\providecommand \bibitemNoStop [0]{.\EOS\space}%
\providecommand \EOS [0]{\spacefactor3000\relax}%
\providecommand \BibitemShut  [1]{\csname bibitem#1\endcsname}%
\let\auto@bib@innerbib\@empty
\bibitem [{\citenamefont {Slutz}\ \emph {et~al.}(2010)\citenamefont {Slutz},
  \citenamefont {Herrmann}, \citenamefont {Vesey}, \citenamefont {Sefkow},
  \citenamefont {Sinars}, \citenamefont {Rovang}, \citenamefont {Peterson},\
  and\ \citenamefont {Cuneo}}]{Slutz2010}%
  \BibitemOpen
  \bibfield  {author} {\bibinfo {author} {\bibfnamefont {S.~A.}\ \bibnamefont
  {Slutz}}, \bibinfo {author} {\bibfnamefont {M.~C.}\ \bibnamefont {Herrmann}},
  \bibinfo {author} {\bibfnamefont {R.~A.}\ \bibnamefont {Vesey}}, \bibinfo
  {author} {\bibfnamefont {A.~B.}\ \bibnamefont {Sefkow}}, \bibinfo {author}
  {\bibfnamefont {D.~B.}\ \bibnamefont {Sinars}}, \bibinfo {author}
  {\bibfnamefont {D.~C.}\ \bibnamefont {Rovang}}, \bibinfo {author}
  {\bibfnamefont {K.~J.}\ \bibnamefont {Peterson}}, \ and\ \bibinfo {author}
  {\bibfnamefont {M.~E.}\ \bibnamefont {Cuneo}},\ }\href {\doibase
  10.1063/1.3333505} {\bibfield  {journal} {\bibinfo  {journal} {Physics of
  Plasmas}\ }\textbf {\bibinfo {volume} {17}} (\bibinfo {year} {2010}),\
  10.1063/1.3333505}\BibitemShut {NoStop}%
\bibitem [{\citenamefont {Awe}\ \emph {et~al.}(2013)\citenamefont {Awe},
  \citenamefont {McBride}, \citenamefont {Jennings}, \citenamefont {Lamppa},
  \citenamefont {Martin}, \citenamefont {Rovang}, \citenamefont {Slutz},
  \citenamefont {Cuneo}, \citenamefont {Owen}, \citenamefont {Sinars},
  \citenamefont {Tomlinson}, \citenamefont {Gomez}, \citenamefont {Hansen},
  \citenamefont {Herrmann}, \citenamefont {McKenney}, \citenamefont {Nakhleh},
  \citenamefont {Robertson}, \citenamefont {Rochau}, \citenamefont {Savage},
  \citenamefont {Schroen},\ and\ \citenamefont {Stygar}}]{Awe2013}%
  \BibitemOpen
  \bibfield  {author} {\bibinfo {author} {\bibfnamefont {T.~J.}\ \bibnamefont
  {Awe}}, \bibinfo {author} {\bibfnamefont {R.~D.}\ \bibnamefont {McBride}},
  \bibinfo {author} {\bibfnamefont {C.~A.}\ \bibnamefont {Jennings}}, \bibinfo
  {author} {\bibfnamefont {D.~C.}\ \bibnamefont {Lamppa}}, \bibinfo {author}
  {\bibfnamefont {M.~R.}\ \bibnamefont {Martin}}, \bibinfo {author}
  {\bibfnamefont {D.~C.}\ \bibnamefont {Rovang}}, \bibinfo {author}
  {\bibfnamefont {S.~A.}\ \bibnamefont {Slutz}}, \bibinfo {author}
  {\bibfnamefont {M.~E.}\ \bibnamefont {Cuneo}}, \bibinfo {author}
  {\bibfnamefont {A.~C.}\ \bibnamefont {Owen}}, \bibinfo {author}
  {\bibfnamefont {D.~B.}\ \bibnamefont {Sinars}}, \bibinfo {author}
  {\bibfnamefont {K.}~\bibnamefont {Tomlinson}}, \bibinfo {author}
  {\bibfnamefont {M.~R.}\ \bibnamefont {Gomez}}, \bibinfo {author}
  {\bibfnamefont {S.~B.}\ \bibnamefont {Hansen}}, \bibinfo {author}
  {\bibfnamefont {M.~C.}\ \bibnamefont {Herrmann}}, \bibinfo {author}
  {\bibfnamefont {J.~L.}\ \bibnamefont {McKenney}}, \bibinfo {author}
  {\bibfnamefont {C.}~\bibnamefont {Nakhleh}}, \bibinfo {author} {\bibfnamefont
  {G.~K.}\ \bibnamefont {Robertson}}, \bibinfo {author} {\bibfnamefont {G.~A.}\
  \bibnamefont {Rochau}}, \bibinfo {author} {\bibfnamefont {M.~E.}\
  \bibnamefont {Savage}}, \bibinfo {author} {\bibfnamefont {D.~G.}\
  \bibnamefont {Schroen}}, \ and\ \bibinfo {author} {\bibfnamefont {W.~A.}\
  \bibnamefont {Stygar}},\ }\href {\doibase 10.1103/PhysRevLett.111.235005}
  {\bibfield  {journal} {\bibinfo  {journal} {Physical Review Letters}\
  }\textbf {\bibinfo {volume} {111}},\ \bibinfo {pages} {1} (\bibinfo {year}
  {2013})}\BibitemShut {NoStop}%
\bibitem [{\citenamefont {Gomez}\ \emph {et~al.}(2014)\citenamefont {Gomez},
  \citenamefont {Slutz}, \citenamefont {Sefkow}, \citenamefont {Sinars},
  \citenamefont {Hahn}, \citenamefont {Hansen}, \citenamefont {Harding},
  \citenamefont {Knapp}, \citenamefont {Schmit}, \citenamefont {Jennings},
  \citenamefont {Awe}, \citenamefont {Geissel}, \citenamefont {Rovang},
  \citenamefont {Chandler}, \citenamefont {Cooper}, \citenamefont {Cuneo},
  \citenamefont {Harvey-Thompson}, \citenamefont {Herrmann}, \citenamefont
  {Hess}, \citenamefont {Johns}, \citenamefont {Lamppa}, \citenamefont
  {Martin}, \citenamefont {McBride}, \citenamefont {Peterson}, \citenamefont
  {Porter}, \citenamefont {Robertson}, \citenamefont {Rochau}, \citenamefont
  {Ruiz}, \citenamefont {Savage}, \citenamefont {Smith}, \citenamefont
  {Stygar},\ and\ \citenamefont {Vesey}}]{Gomez2014}%
  \BibitemOpen
  \bibfield  {author} {\bibinfo {author} {\bibfnamefont {M.~R.}\ \bibnamefont
  {Gomez}}, \bibinfo {author} {\bibfnamefont {S.~A.}\ \bibnamefont {Slutz}},
  \bibinfo {author} {\bibfnamefont {A.~B.}\ \bibnamefont {Sefkow}}, \bibinfo
  {author} {\bibfnamefont {D.~B.}\ \bibnamefont {Sinars}}, \bibinfo {author}
  {\bibfnamefont {K.~D.}\ \bibnamefont {Hahn}}, \bibinfo {author}
  {\bibfnamefont {S.~B.}\ \bibnamefont {Hansen}}, \bibinfo {author}
  {\bibfnamefont {E.~C.}\ \bibnamefont {Harding}}, \bibinfo {author}
  {\bibfnamefont {P.~F.}\ \bibnamefont {Knapp}}, \bibinfo {author}
  {\bibfnamefont {P.~F.}\ \bibnamefont {Schmit}}, \bibinfo {author}
  {\bibfnamefont {C.~A.}\ \bibnamefont {Jennings}}, \bibinfo {author}
  {\bibfnamefont {T.~J.}\ \bibnamefont {Awe}}, \bibinfo {author} {\bibfnamefont
  {M.}~\bibnamefont {Geissel}}, \bibinfo {author} {\bibfnamefont {D.~C.}\
  \bibnamefont {Rovang}}, \bibinfo {author} {\bibfnamefont {G.~A.}\
  \bibnamefont {Chandler}}, \bibinfo {author} {\bibfnamefont {G.~W.}\
  \bibnamefont {Cooper}}, \bibinfo {author} {\bibfnamefont {M.~E.}\
  \bibnamefont {Cuneo}}, \bibinfo {author} {\bibfnamefont {A.~J.}\ \bibnamefont
  {Harvey-Thompson}}, \bibinfo {author} {\bibfnamefont {M.~C.}\ \bibnamefont
  {Herrmann}}, \bibinfo {author} {\bibfnamefont {M.~H.}\ \bibnamefont {Hess}},
  \bibinfo {author} {\bibfnamefont {O.}~\bibnamefont {Johns}}, \bibinfo
  {author} {\bibfnamefont {D.~C.}\ \bibnamefont {Lamppa}}, \bibinfo {author}
  {\bibfnamefont {M.~R.}\ \bibnamefont {Martin}}, \bibinfo {author}
  {\bibfnamefont {R.~D.}\ \bibnamefont {McBride}}, \bibinfo {author}
  {\bibfnamefont {K.~J.}\ \bibnamefont {Peterson}}, \bibinfo {author}
  {\bibfnamefont {J.~L.}\ \bibnamefont {Porter}}, \bibinfo {author}
  {\bibfnamefont {G.~K.}\ \bibnamefont {Robertson}}, \bibinfo {author}
  {\bibfnamefont {G.~A.}\ \bibnamefont {Rochau}}, \bibinfo {author}
  {\bibfnamefont {C.~L.}\ \bibnamefont {Ruiz}}, \bibinfo {author}
  {\bibfnamefont {M.~E.}\ \bibnamefont {Savage}}, \bibinfo {author}
  {\bibfnamefont {I.~C.}\ \bibnamefont {Smith}}, \bibinfo {author}
  {\bibfnamefont {W.~A.}\ \bibnamefont {Stygar}}, \ and\ \bibinfo {author}
  {\bibfnamefont {R.~A.}\ \bibnamefont {Vesey}},\ }\href {\doibase
  10.1103/PhysRevLett.113.155003} {\bibfield  {journal} {\bibinfo  {journal}
  {Physical Review Letters}\ }\textbf {\bibinfo {volume} {113}},\ \bibinfo
  {pages} {1} (\bibinfo {year} {2014})}\BibitemShut {NoStop}%
\bibitem [{\citenamefont {McBride}\ \emph {et~al.}(2012)\citenamefont
  {McBride}, \citenamefont {Slutz}, \citenamefont {Jennings}, \citenamefont
  {Sinars}, \citenamefont {Cuneo}, \citenamefont {Herrmann}, \citenamefont
  {Lemke}, \citenamefont {Martin}, \citenamefont {Vesey}, \citenamefont
  {Peterson}, \citenamefont {Sefkow}, \citenamefont {Nakhleh}, \citenamefont
  {Blue}, \citenamefont {Killebrew}, \citenamefont {Schroen}, \citenamefont
  {Rogers}, \citenamefont {Laspe}, \citenamefont {Lopez}, \citenamefont
  {Smith}, \citenamefont {Atherton}, \citenamefont {Savage}, \citenamefont
  {Stygar},\ and\ \citenamefont {Porter}}]{McBride2012}%
  \BibitemOpen
  \bibfield  {author} {\bibinfo {author} {\bibfnamefont {R.~D.}\ \bibnamefont
  {McBride}}, \bibinfo {author} {\bibfnamefont {S.~A.}\ \bibnamefont {Slutz}},
  \bibinfo {author} {\bibfnamefont {C.~A.}\ \bibnamefont {Jennings}}, \bibinfo
  {author} {\bibfnamefont {D.~B.}\ \bibnamefont {Sinars}}, \bibinfo {author}
  {\bibfnamefont {M.~E.}\ \bibnamefont {Cuneo}}, \bibinfo {author}
  {\bibfnamefont {M.~C.}\ \bibnamefont {Herrmann}}, \bibinfo {author}
  {\bibfnamefont {R.~W.}\ \bibnamefont {Lemke}}, \bibinfo {author}
  {\bibfnamefont {M.~R.}\ \bibnamefont {Martin}}, \bibinfo {author}
  {\bibfnamefont {R.~A.}\ \bibnamefont {Vesey}}, \bibinfo {author}
  {\bibfnamefont {K.~J.}\ \bibnamefont {Peterson}}, \bibinfo {author}
  {\bibfnamefont {A.~B.}\ \bibnamefont {Sefkow}}, \bibinfo {author}
  {\bibfnamefont {C.}~\bibnamefont {Nakhleh}}, \bibinfo {author} {\bibfnamefont
  {B.~E.}\ \bibnamefont {Blue}}, \bibinfo {author} {\bibfnamefont
  {K.}~\bibnamefont {Killebrew}}, \bibinfo {author} {\bibfnamefont
  {D.}~\bibnamefont {Schroen}}, \bibinfo {author} {\bibfnamefont {T.~J.}\
  \bibnamefont {Rogers}}, \bibinfo {author} {\bibfnamefont {A.}~\bibnamefont
  {Laspe}}, \bibinfo {author} {\bibfnamefont {M.~R.}\ \bibnamefont {Lopez}},
  \bibinfo {author} {\bibfnamefont {I.~C.}\ \bibnamefont {Smith}}, \bibinfo
  {author} {\bibfnamefont {B.~W.}\ \bibnamefont {Atherton}}, \bibinfo {author}
  {\bibfnamefont {M.}~\bibnamefont {Savage}}, \bibinfo {author} {\bibfnamefont
  {W.~A.}\ \bibnamefont {Stygar}}, \ and\ \bibinfo {author} {\bibfnamefont
  {J.~L.}\ \bibnamefont {Porter}},\ }\href {\doibase
  10.1103/PhysRevLett.109.135004} {\bibfield  {journal} {\bibinfo  {journal}
  {Phys. Rev. Lett.}\ }\textbf {\bibinfo {volume} {109}},\ \bibinfo {pages}
  {135004} (\bibinfo {year} {2012})}\BibitemShut {NoStop}%
\bibitem [{\citenamefont {Seyler}, \citenamefont {Martin},\ and\ \citenamefont
  {Hamlin}(2018)}]{Seyler2018}%
  \BibitemOpen
  \bibfield  {author} {\bibinfo {author} {\bibfnamefont {C.~E.}\ \bibnamefont
  {Seyler}}, \bibinfo {author} {\bibfnamefont {M.~R.}\ \bibnamefont {Martin}},
  \ and\ \bibinfo {author} {\bibfnamefont {N.~D.}\ \bibnamefont {Hamlin}},\
  }\href {\doibase 10.1063/1.5028365} {\bibfield  {journal} {\bibinfo
  {journal} {Physics of Plasmas}\ }\textbf {\bibinfo {volume} {25}},\ \bibinfo
  {pages} {062711} (\bibinfo {year} {2018})},\ \Eprint
  {http://arxiv.org/abs/https://doi.org/10.1063/1.5028365}
  {https://doi.org/10.1063/1.5028365} \BibitemShut {NoStop}%
\bibitem [{\citenamefont {Ampleford}\ \emph {et~al.}(2019)\citenamefont
  {Ampleford}, \citenamefont {Jennings}, \citenamefont {Harding}, \citenamefont
  {Gomez}, \citenamefont {Webb}, \citenamefont {Schmit}, \citenamefont {Awe},
  \citenamefont {Knapp}, \citenamefont {Harvey-Thompson}, \citenamefont {Weis},
  \citenamefont {Slutz}, \citenamefont {Hansen}, \citenamefont {Chandler},
  \citenamefont {Dunham}, \citenamefont {Geissel}, \citenamefont {Fisher},
  \citenamefont {adn M.~E.~Glinsky}, \citenamefont {Hahn}, \citenamefont
  {Lamppa}, \citenamefont {Lucero}, \citenamefont {Mangan}, \citenamefont
  {Martin}, \citenamefont {Moore}, \citenamefont {Paguio}, \citenamefont
  {Perea}, \citenamefont {Robertson}, \citenamefont {Ruiz}, \citenamefont
  {Smith}, \citenamefont {Smith}, \citenamefont {Speas}, \citenamefont
  {Whittemore}, \citenamefont {Yu}, \citenamefont {McBride}, \citenamefont
  {Jones}, \citenamefont {Peterson}, \citenamefont {Rochau},\ and\
  \citenamefont {Sinars}}]{Ampleford2019}%
  \BibitemOpen
  \bibfield  {author} {\bibinfo {author} {\bibfnamefont {D.~J.}\ \bibnamefont
  {Ampleford}}, \bibinfo {author} {\bibfnamefont {C.~A.}\ \bibnamefont
  {Jennings}}, \bibinfo {author} {\bibfnamefont {E.~C.}\ \bibnamefont
  {Harding}}, \bibinfo {author} {\bibfnamefont {M.~R.}\ \bibnamefont {Gomez}},
  \bibinfo {author} {\bibfnamefont {T.}~\bibnamefont {Webb}}, \bibinfo {author}
  {\bibfnamefont {P.}~\bibnamefont {Schmit}}, \bibinfo {author} {\bibfnamefont
  {T.~J.}\ \bibnamefont {Awe}}, \bibinfo {author} {\bibfnamefont {P.~F.}\
  \bibnamefont {Knapp}}, \bibinfo {author} {\bibfnamefont {A.~J.}\ \bibnamefont
  {Harvey-Thompson}}, \bibinfo {author} {\bibfnamefont {M.~R.}\ \bibnamefont
  {Weis}}, \bibinfo {author} {\bibfnamefont {S.~A.}\ \bibnamefont {Slutz}},
  \bibinfo {author} {\bibfnamefont {S.~B.}\ \bibnamefont {Hansen}}, \bibinfo
  {author} {\bibfnamefont {G.~A.}\ \bibnamefont {Chandler}}, \bibinfo {author}
  {\bibfnamefont {G.}~\bibnamefont {Dunham}}, \bibinfo {author} {\bibfnamefont
  {M.}~\bibnamefont {Geissel}}, \bibinfo {author} {\bibfnamefont
  {J.}~\bibnamefont {Fisher}}, \bibinfo {author} {\bibfnamefont {D.~F.}\
  \bibnamefont {adn M.~E.~Glinsky}}, \bibinfo {author} {\bibfnamefont {K.~D.}\
  \bibnamefont {Hahn}}, \bibinfo {author} {\bibfnamefont {D.}~\bibnamefont
  {Lamppa}}, \bibinfo {author} {\bibfnamefont {L.}~\bibnamefont {Lucero}},
  \bibinfo {author} {\bibfnamefont {M.}~\bibnamefont {Mangan}}, \bibinfo
  {author} {\bibfnamefont {M.}~\bibnamefont {Martin}}, \bibinfo {author}
  {\bibfnamefont {T.}~\bibnamefont {Moore}}, \bibinfo {author} {\bibfnamefont
  {R.}~\bibnamefont {Paguio}}, \bibinfo {author} {\bibfnamefont
  {L.}~\bibnamefont {Perea}}, \bibinfo {author} {\bibfnamefont
  {G.}~\bibnamefont {Robertson}}, \bibinfo {author} {\bibfnamefont
  {C.}~\bibnamefont {Ruiz}}, \bibinfo {author} {\bibfnamefont {G.~E.}\
  \bibnamefont {Smith}}, \bibinfo {author} {\bibfnamefont {I.~C.}\ \bibnamefont
  {Smith}}, \bibinfo {author} {\bibfnamefont {C.~S.}\ \bibnamefont {Speas}},
  \bibinfo {author} {\bibfnamefont {K.}~\bibnamefont {Whittemore}}, \bibinfo
  {author} {\bibfnamefont {E.}~\bibnamefont {Yu}}, \bibinfo {author}
  {\bibfnamefont {R.}~\bibnamefont {McBride}}, \bibinfo {author} {\bibfnamefont
  {B.}~\bibnamefont {Jones}}, \bibinfo {author} {\bibfnamefont {K.~J.}\
  \bibnamefont {Peterson}}, \bibinfo {author} {\bibfnamefont {G.~A.}\
  \bibnamefont {Rochau}}, \ and\ \bibinfo {author} {\bibfnamefont {D.~B.}\
  \bibnamefont {Sinars}},\ }\href@noop {} {\enquote {\bibinfo {title} {Improved
  morphology and reproducibility of magnetized liner inertial fusion
  experiments},}\ } (\bibinfo {year} {2019}),\ \bibinfo {note} {in preparation
  for \textit{Physics of Plasmas}}\BibitemShut {NoStop}%
\bibitem [{\citenamefont {Peterson}\ \emph {et~al.}(2012)\citenamefont
  {Peterson}, \citenamefont {Sinars}, \citenamefont {Yu}, \citenamefont
  {Herrmann}, \citenamefont {Cuneo}, \citenamefont {Slutz}, \citenamefont
  {Smith}, \citenamefont {Atherton}, \citenamefont {Knudson},\ and\
  \citenamefont {Nakhleh}}]{peterson2012electrothermal}%
  \BibitemOpen
  \bibfield  {author} {\bibinfo {author} {\bibfnamefont {K.~J.}\ \bibnamefont
  {Peterson}}, \bibinfo {author} {\bibfnamefont {D.~B.}\ \bibnamefont
  {Sinars}}, \bibinfo {author} {\bibfnamefont {E.~P.}\ \bibnamefont {Yu}},
  \bibinfo {author} {\bibfnamefont {M.~C.}\ \bibnamefont {Herrmann}}, \bibinfo
  {author} {\bibfnamefont {M.~E.}\ \bibnamefont {Cuneo}}, \bibinfo {author}
  {\bibfnamefont {S.~A.}\ \bibnamefont {Slutz}}, \bibinfo {author}
  {\bibfnamefont {I.~C.}\ \bibnamefont {Smith}}, \bibinfo {author}
  {\bibfnamefont {B.~W.}\ \bibnamefont {Atherton}}, \bibinfo {author}
  {\bibfnamefont {M.~D.}\ \bibnamefont {Knudson}}, \ and\ \bibinfo {author}
  {\bibfnamefont {C.}~\bibnamefont {Nakhleh}},\ }\href@noop {} {\bibfield
  {journal} {\bibinfo  {journal} {Physics of Plasmas}\ }\textbf {\bibinfo
  {volume} {19}},\ \bibinfo {pages} {092701} (\bibinfo {year}
  {2012})}\BibitemShut {NoStop}%
\bibitem [{\citenamefont {Peterson}\ \emph {et~al.}(2014)\citenamefont
  {Peterson}, \citenamefont {Awe}, \citenamefont {Edmund}, \citenamefont
  {Sinars}, \citenamefont {Field}, \citenamefont {Cuneo}, \citenamefont
  {Herrmann}, \citenamefont {Savage}, \citenamefont {Schroen}, \citenamefont
  {Tomlinson} \emph {et~al.}}]{peterson2014electrothermal}%
  \BibitemOpen
  \bibfield  {author} {\bibinfo {author} {\bibfnamefont {K.~J.}\ \bibnamefont
  {Peterson}}, \bibinfo {author} {\bibfnamefont {T.~J.}\ \bibnamefont {Awe}},
  \bibinfo {author} {\bibfnamefont {P.~Y.}\ \bibnamefont {Edmund}}, \bibinfo
  {author} {\bibfnamefont {D.~B.}\ \bibnamefont {Sinars}}, \bibinfo {author}
  {\bibfnamefont {E.~S.}\ \bibnamefont {Field}}, \bibinfo {author}
  {\bibfnamefont {M.~E.}\ \bibnamefont {Cuneo}}, \bibinfo {author}
  {\bibfnamefont {M.~C.}\ \bibnamefont {Herrmann}}, \bibinfo {author}
  {\bibfnamefont {M.}~\bibnamefont {Savage}}, \bibinfo {author} {\bibfnamefont
  {D.}~\bibnamefont {Schroen}}, \bibinfo {author} {\bibfnamefont
  {K.}~\bibnamefont {Tomlinson}},  \emph {et~al.},\ }\href@noop {} {\bibfield
  {journal} {\bibinfo  {journal} {Physical Review Letters}\ }\textbf {\bibinfo
  {volume} {112}},\ \bibinfo {pages} {135002} (\bibinfo {year}
  {2014})}\BibitemShut {NoStop}%
\bibitem [{\citenamefont {Perez}\ and\ \citenamefont
  {Boldyrev}(2009)}]{perez2009role}%
  \BibitemOpen
  \bibfield  {author} {\bibinfo {author} {\bibfnamefont {J.~C.}\ \bibnamefont
  {Perez}}\ and\ \bibinfo {author} {\bibfnamefont {S.}~\bibnamefont
  {Boldyrev}},\ }\href@noop {} {\bibfield  {journal} {\bibinfo  {journal}
  {Physical Review Letters}\ }\textbf {\bibinfo {volume} {102}},\ \bibinfo
  {pages} {025003} (\bibinfo {year} {2009})}\BibitemShut {NoStop}%
\bibitem [{\citenamefont {Glinsky}\ and\ \citenamefont
  {Hjorth}(2019)}]{glinsky2019helicity}%
  \BibitemOpen
  \bibfield  {author} {\bibinfo {author} {\bibfnamefont {M.~E.}\ \bibnamefont
  {Glinsky}}\ and\ \bibinfo {author} {\bibfnamefont {P.~G.}\ \bibnamefont
  {Hjorth}},\ }\href {https://arxiv.org/abs/1912.04895} {\enquote {\bibinfo
  {title} {Helicity in Hamiltonian dynamical systems},}\ }\bibinfo {type}
  {Tech. Rep.}\ \bibinfo {number} {SAND2019-14731}\ (\bibinfo  {institution}
  {Sandia National Laboratories},\ \bibinfo {year} {2019})\ \bibinfo {note}
  {arXiv:1912.04895}\BibitemShut {NoStop}%
\bibitem [{\citenamefont {Taylor}(1986)}]{taylor1986relaxation}%
  \BibitemOpen
  \bibfield  {author} {\bibinfo {author} {\bibfnamefont {J.}~\bibnamefont
  {Taylor}},\ }\href@noop {} {\bibfield  {journal} {\bibinfo  {journal}
  {Reviews of Modern Physics}\ }\textbf {\bibinfo {volume} {58}},\ \bibinfo
  {pages} {741} (\bibinfo {year} {1986})}\BibitemShut {NoStop}%
\bibitem [{\citenamefont {Yager-Elorriaga}\ \emph {et~al.}(2018)\citenamefont
  {Yager-Elorriaga}, \citenamefont {Lau}, \citenamefont {Zhang}, \citenamefont
  {Campbell}, \citenamefont {Steiner}, \citenamefont {Jordan}, \citenamefont
  {McBride},\ and\ \citenamefont {Gilgenbach}}]{yager2018}%
  \BibitemOpen
  \bibfield  {author} {\bibinfo {author} {\bibfnamefont {D.~A.}\ \bibnamefont
  {Yager-Elorriaga}}, \bibinfo {author} {\bibfnamefont {Y.}~\bibnamefont
  {Lau}}, \bibinfo {author} {\bibfnamefont {P.}~\bibnamefont {Zhang}}, \bibinfo
  {author} {\bibfnamefont {P.~C.}\ \bibnamefont {Campbell}}, \bibinfo {author}
  {\bibfnamefont {A.~M.}\ \bibnamefont {Steiner}}, \bibinfo {author}
  {\bibfnamefont {N.~M.}\ \bibnamefont {Jordan}}, \bibinfo {author}
  {\bibfnamefont {R.~D.}\ \bibnamefont {McBride}}, \ and\ \bibinfo {author}
  {\bibfnamefont {R.~M.}\ \bibnamefont {Gilgenbach}},\ }\href@noop {}
  {\bibfield  {journal} {\bibinfo  {journal} {Physics of Plasmas}\ }\textbf
  {\bibinfo {volume} {25}},\ \bibinfo {pages} {056307} (\bibinfo {year}
  {2018})}\BibitemShut {NoStop}%
\bibitem [{\citenamefont {Mallat}(2012)}]{Mallat2012}%
  \BibitemOpen
  \bibfield  {author} {\bibinfo {author} {\bibfnamefont {S.}~\bibnamefont
  {Mallat}},\ }\href@noop {} {\bibfield  {journal} {\bibinfo  {journal}
  {Communications on Pure and Applied Mathematics}\ }\textbf {\bibinfo {volume}
  {65}},\ \bibinfo {pages} {1331} (\bibinfo {year} {2012})}\BibitemShut
  {NoStop}%
\bibitem [{\citenamefont {Bruna}\ and\ \citenamefont
  {Mallat}(2013)}]{Bruna2013}%
  \BibitemOpen
  \bibfield  {author} {\bibinfo {author} {\bibfnamefont {J.}~\bibnamefont
  {Bruna}}\ and\ \bibinfo {author} {\bibfnamefont {S.}~\bibnamefont {Mallat}},\
  }\href {\doibase 10.1109/TPAMI.2012.230} {\bibfield  {journal} {\bibinfo
  {journal} {IEEE Transactions on Pattern Analysis and Machine Intelligence}\
  }\textbf {\bibinfo {volume} {35}},\ \bibinfo {pages} {1872} (\bibinfo {year}
  {2013})},\ \Eprint {http://arxiv.org/abs/1203.1513} {arXiv:1203.1513}
  \BibitemShut {NoStop}%
\bibitem [{\citenamefont {le~Cun}(1989)}]{LeCun1989}%
  \BibitemOpen
  \bibfield  {author} {\bibinfo {author} {\bibfnamefont {Y.}~\bibnamefont
  {Le~Cun}},\ }\href@noop {} {\enquote {\bibinfo {title} {{Generalization and
  network design strategies. Technical Report CRG-TR-89-4}},}\ }\bibinfo {type}
  {Tech. Rep.}\ (\bibinfo  {institution} {University of Toronto, Department of
  Computer Science},\ \bibinfo {year} {1989})\BibitemShut {NoStop}%
\bibitem [{\citenamefont {Ning}\ \emph {et~al.}(2005)\citenamefont {Ning},
  \citenamefont {Delhomme}, \citenamefont {le~Cun}, \citenamefont {Piano},
  \citenamefont {Bottou},\ and\ \citenamefont {Barbano}}]{Ning2005}%
  \BibitemOpen
  \bibfield  {author} {\bibinfo {author} {\bibfnamefont {F.}~\bibnamefont
  {Ning}}, \bibinfo {author} {\bibfnamefont {D.}~\bibnamefont {Delhomme}},
  \bibinfo {author} {\bibfnamefont {Y.}~\bibnamefont {le~Cun}}, \bibinfo
  {author} {\bibfnamefont {F.}~\bibnamefont {Piano}}, \bibinfo {author}
  {\bibfnamefont {L.}~\bibnamefont {Bottou}}, \ and\ \bibinfo {author}
  {\bibfnamefont {P.}~\bibnamefont {Barbano}},\ }\href@noop {} {\bibfield
  {journal} {\bibinfo  {journal} {IEEE Transactions on Image Processing}\
  }\textbf {\bibinfo {volume} {14}},\ \bibinfo {pages} {1360} (\bibinfo {year}
  {2005})}\BibitemShut {NoStop}%
\bibitem [{\citenamefont {Goodfellow}\ \emph {et~al.}(2014)\citenamefont
  {Goodfellow}, \citenamefont {Pouget-Abadie}, \citenamefont {Mirza},
  \citenamefont {Xu}, \citenamefont {Warde-Farley}, \citenamefont {Ozair},
  \citenamefont {Courville},\ and\ \citenamefont {Bengio}}]{Goodfellow2014}%
  \BibitemOpen
  \bibfield  {author} {\bibinfo {author} {\bibfnamefont {I.}~\bibnamefont
  {Goodfellow}}, \bibinfo {author} {\bibfnamefont {J.}~\bibnamefont
  {Pouget-Abadie}}, \bibinfo {author} {\bibfnamefont {M.}~\bibnamefont
  {Mirza}}, \bibinfo {author} {\bibfnamefont {B.}~\bibnamefont {Xu}}, \bibinfo
  {author} {\bibfnamefont {D.}~\bibnamefont {Warde-Farley}}, \bibinfo {author}
  {\bibfnamefont {S.}~\bibnamefont {Ozair}}, \bibinfo {author} {\bibfnamefont
  {A.}~\bibnamefont {Courville}}, \ and\ \bibinfo {author} {\bibfnamefont
  {Y.}~\bibnamefont {Bengio}},\ }in\ \href@noop {} {\emph {\bibinfo {booktitle}
  {Advances in Neural Information Processing Systems}}}\ (\bibinfo {year}
  {2014})\ pp.\ \bibinfo {pages} {2672--2680}\BibitemShut {NoStop}%
\bibitem [{\citenamefont {LeCun}, \citenamefont {Kavukcuoglu},\ and\
  \citenamefont {Farabet}(2010)}]{LeCun2010}%
  \BibitemOpen
  \bibfield  {author} {\bibinfo {author} {\bibfnamefont {Y.}~\bibnamefont
  {LeCun}}, \bibinfo {author} {\bibfnamefont {K.}~\bibnamefont {Kavukcuoglu}},
  \ and\ \bibinfo {author} {\bibfnamefont {C.}~\bibnamefont {Farabet}},\ }in\
  \href {\doibase 10.1109/ISCAS.2010.5537907} {\emph {\bibinfo {booktitle}
  {International Symposium on Circuits and Systems {(ISCAS} 2010), May 30 -
  June 2, 2010, Paris, France}}}\ (\bibinfo {year} {2010})\ pp.\ \bibinfo
  {pages} {253--256}\BibitemShut {NoStop}%
\bibitem [{\citenamefont {Goodfellow}, \citenamefont {Bengio},\ and\
  \citenamefont {Courville}(2016)}]{goodfellow2016deep}%
  \BibitemOpen
  \bibfield  {author} {\bibinfo {author} {\bibfnamefont {I.}~\bibnamefont
  {Goodfellow}}, \bibinfo {author} {\bibfnamefont {Y.}~\bibnamefont {Bengio}},
  \ and\ \bibinfo {author} {\bibfnamefont {A.}~\bibnamefont {Courville}},\
  }\href@noop {} {\emph {\bibinfo {title} {Deep Learning}}},\ \bibinfo {number}
  {Chap. 15, Sec. 2}\ (\bibinfo  {publisher} {MIT press},\ \bibinfo {year}
  {2016})\BibitemShut {NoStop}%
\bibitem [{\citenamefont {Mallat}(1999)}]{mallat1999wavelet}%
  \BibitemOpen
  \bibfield  {author} {\bibinfo {author} {\bibfnamefont {S.}~\bibnamefont
  {Mallat}},\ }\href@noop {} {\emph {\bibinfo {title} {A Wavelet Tour of Signal
  Processing}}}\ (\bibinfo  {publisher} {Elsevier},\ \bibinfo {year}
  {1999})\BibitemShut {NoStop}%
\bibitem [{\citenamefont {Krommes}(2002)}]{krommes2002}%
  \BibitemOpen
  \bibfield  {author} {\bibinfo {author} {\bibfnamefont {J.~A.}\ \bibnamefont
  {Krommes}},\ }\href@noop {} {\bibfield  {journal} {\bibinfo  {journal}
  {Physics Reports}\ }\textbf {\bibinfo {volume} {360}},\ \bibinfo {pages} {1}
  (\bibinfo {year} {2002})}\BibitemShut {NoStop}%
\bibitem [{\citenamefont {Glinsky}\ \emph {et~al.}(2019)\citenamefont
  {Glinsky}, \citenamefont {Moore}, \citenamefont {Lewis}, \citenamefont
  {Weis}, \citenamefont {Jennings}, \citenamefont {Ampleford}, \citenamefont
  {Harding}, \citenamefont {Knapp}, \citenamefont {Gomez},\ and\ \citenamefont
  {Lussiez}}]{glinsky.et.al.19}%
  \BibitemOpen
  \bibfield  {author} {\bibinfo {author} {\bibfnamefont {M.~E.}\ \bibnamefont
  {Glinsky}}, \bibinfo {author} {\bibfnamefont {T.~W.}\ \bibnamefont {Moore}},
  \bibinfo {author} {\bibfnamefont {W.~E.}\ \bibnamefont {Lewis}}, \bibinfo
  {author} {\bibfnamefont {M.~R.}\ \bibnamefont {Weis}}, \bibinfo {author}
  {\bibfnamefont {C.~A.}\ \bibnamefont {Jennings}}, \bibinfo {author}
  {\bibfnamefont {D.~A.}\ \bibnamefont {Ampleford}}, \bibinfo {author}
  {\bibfnamefont {E.~C.}\ \bibnamefont {Harding}}, \bibinfo {author}
  {\bibfnamefont {P.~F.}\ \bibnamefont {Knapp}}, \bibinfo {author}
  {\bibfnamefont {M.~R.}\ \bibnamefont {Gomez}}, \ and\ \bibinfo {author}
  {\bibfnamefont {S.~E.}\ \bibnamefont {Lussiez}},\ }\href
  {http://arxiv.org/abs/1911.02359} {\enquote {\bibinfo {title} {Quantification
  of MagLIF morphology using the Mallat Scattering Transformation},}\ }\bibinfo
  {type} {Tech. Rep.}\ \bibinfo {number} {SAND2019-11910}\ (\bibinfo
  {institution} {Sandia National Laboratories},\ \bibinfo {year} {2019})\
  \bibinfo {note} {arXiv:1911.02359}\BibitemShut {NoStop}%
\bibitem [{\citenamefont {Glinsky}(2011)}]{glinsky.11}%
  \BibitemOpen
  \bibfield  {author} {\bibinfo {author} {\bibfnamefont {M.~E.}\ \bibnamefont
  {Glinsky}},\ }\href {http://arxiv.org/abs/1106.4369} {\enquote {\bibinfo
  {title} {A new perspective on renormalization: the scattering
  transformation},}\ }\bibinfo {type} {Tech. Rep.}\ (\bibinfo  {institution}
  {CSIRO},\ \bibinfo {year} {2011})\ \bibinfo {note}
  {arXiv:1106.4369}\BibitemShut {NoStop}%
\bibitem [{\citenamefont {Bishop}(2006)}]{bishop}%
  \BibitemOpen
  \bibfield  {author} {\bibinfo {author} {\bibfnamefont {C.~M.}\ \bibnamefont
  {Bishop}},\ }\href@noop {} {\emph {\bibinfo {title} {Pattern Recognition and
  Machine Learning}}}\ (\bibinfo  {publisher} {Springer},\ \bibinfo {year}
  {2006})\BibitemShut {NoStop}%
\bibitem [{\citenamefont {Chittenden}\ \emph {et~al.}(2004)\citenamefont
  {Chittenden}, \citenamefont {Lebedev}, \citenamefont {Jennings},
  \citenamefont {Bland},\ and\ \citenamefont {Ciardi}}]{Chittenden2004}%
  \BibitemOpen
  \bibfield  {author} {\bibinfo {author} {\bibfnamefont {J.~P.}\ \bibnamefont
  {Chittenden}}, \bibinfo {author} {\bibfnamefont {S.~V.}\ \bibnamefont
  {Lebedev}}, \bibinfo {author} {\bibfnamefont {C.~A.}\ \bibnamefont
  {Jennings}}, \bibinfo {author} {\bibfnamefont {S.~N.}\ \bibnamefont {Bland}},
  \ and\ \bibinfo {author} {\bibfnamefont {A.}~\bibnamefont {Ciardi}},\ }\href
  {\doibase 10.1088/0741-3335/46/12B/039} {\bibfield  {journal} {\bibinfo
  {journal} {Plasma Physics and Controlled Fusion}\ }\textbf {\bibinfo {volume}
  {46}} (\bibinfo {year} {2004}),\ 10.1088/0741-3335/46/12B/039}\BibitemShut
  {NoStop}%
\bibitem [{\citenamefont {Atiyah}\ and\ \citenamefont
  {Singer}(1963)}]{atiyah1963index}%
  \BibitemOpen
  \bibfield  {author} {\bibinfo {author} {\bibfnamefont {M.~F.}\ \bibnamefont
  {Atiyah}}\ and\ \bibinfo {author} {\bibfnamefont {I.~M.}\ \bibnamefont
  {Singer}},\ }\href@noop {} {\bibfield  {journal} {\bibinfo  {journal}
  {Bulletin of the American Mathematical Society}\ }\textbf {\bibinfo {volume}
  {69}},\ \bibinfo {pages} {422} (\bibinfo {year} {1963})}\BibitemShut
  {NoStop}%
\end{thebibliography}

%

\end{document}